\documentclass[%
 reprint,amsmath,amssymb,aps]{revtex4-2}
\usepackage{physics}
\usepackage{subfigure}
\usepackage{graphicx}
\usepackage{xcolor}
\usepackage{bm}
\usepackage{hyperref}
\usepackage{times}
\usepackage{soul}

\usepackage{tabularx}  
\usepackage{makecell}  

    \def \a{\alpha}   \def \b{\beta}

   \def \d{\delta}

\def \>{\rangle} 
\def \<{\langle} 
 
\def\be{\begin{equation}} 
\def\ee{\end{equation}} 
\def\longrightharpoonup{\relbar\joinrel\rightharpoonup}
\def\longleftharpoondown{\leftharpoondown\joinrel\relbar}

\def\longrightleftharpoons{
  \mathop{
    \vcenter{
      \hbox{
      \ooalign{
        \raise1pt\hbox{$\longrightharpoonup\joinrel$}\crcr
    \lower1pt\hbox{$\longleftharpoondown\joinrel$}
    }
      }
    }
  }
}

\newcommand \bea {\begin{eqnarray}} 
\newcommand \eea {\end{eqnarray}}

\newcommand{\comment}[1]{{\color{black}#1}}

\begin{document}
\title{A universal niche geometry governs the response of ecosystems to environmental perturbations}
\author{Akshit Goyal$^{1,2}$} 
\email{akshitg@icts.res.in}
\author{Jason W. Rocks$^3$} 
\author{Pankaj Mehta$^{3,4}$}
\email{pankajm@bu.edu}
\address{$^1$ Department of Physics, Massachusetts Insitute of Technology, Cambridge, MA 02139.}
\address{$^2$ International Centre for Theoretical Sciences, Tata Institute of Fundamental Research, Bengaluru 560089.}
\address{$^3$ Department of Physics, Boston University, Boston, MA 02215.}
\address{$^4$ Faculty of Computing and Data Sciences, Boston University, Boston, MA 02215.}

\begin{abstract}
How ecosystems respond to environmental perturbations is a fundamental question in ecology, made especially challenging due to the strong coupling between species and their environment. \comment{ Here, we introduce a theoretical framework for calculating the steady-state response of ecosystems to environmental perturbations in generalized consumer-resource}. Our construction is applicable to a wide class of systems, including models with non-reciprocal interactions, cross-feeding, and non-linear growth/consumption rates. Within our framework, all ecological variables are embedded into four distinct vector spaces and ecological interactions are represented by geometric transformations between these spaces. We show that near a steady state, such geometric transformations directly map environmental perturbations -- in resource availability and mortality rates -- to shifts in niche structure. We illustrate these ideas in a variety of settings including a minimal model for pH-induced toxicity in bacterial denitrification. \comment{We end by discussing the biological implications of our framework. In particular, we show that it is extremely difficult to distinguish cooperative and competitive interactions by measuring species' responses to external perturbations.}
\end{abstract}

\maketitle

One of the most challenging and interesting problems in theoretical ecology is to develop methods for understanding how ecosystems respond to environmental perturbations~\cite{ives2007stability,mccann2000diversity}. This problem is especially acute in diverse ecosystems where species and environment are often strongly coupled due to environmental feedbacks. Characterizing how ecosystems respond to changes in their environment has important practical and theoretical consequences. Practically, understanding how ecosystems respond to changes in resource concentrations is a fundamental concern of environmental science and human health and disease  \cite{lozupone2012diversity,sommer2017resilience} . For example, how does perturbing the nitrification process affect water quality~\cite{chen2006nitrification}? How do changes in diet affect the composition and function of the human gut microbiome~\cite{zhang2012structural,relman2012human}?

From a theoretical vantage point, we know that one of the most successful ways to investigate physical systems is to analyze how they respond to perturbations~\cite{marconi2008fluctuation,bender2013advanced}. This is true for both classical and quantum systems, where objects like static and dynamic susceptibilities are a staple of theoretical treatments of subjects as diverse as electromagnetism, thermodynamics,  and quantum field theory~\cite{kubo1957statistical,schubert2001perturbative}. \comment {For this reason, the theory of linear response has become a pillar of modern theoretical physics. In the context of ecology, techniques for characterizing ecological responses have been largely limited to \comment{asking how ecosystems respond to changes} in the abundances of species or resources~\cite{may1972will,allesina2012stability,butler2018stability}. However, we still lack a framework for understanding how ecosystems respond to environmental perturbations such as changes in resource supplies or the mortality rates. From a mathematical point of view, this amounts to asking how dynamical variables (i.e., species and resource abundances)  respond to small changes in external parameters characterizing the environment (i.e., resource supply rates and mortality rates). }

Here, we address this challenge in the context of generalized consumer-resource models, one of the most widely studied and influential theoretical models in ecology~\cite{chesson1990macarthur,tilman1982resource}. Consumer-resource models (CRMs) have played a central role in the development of modern ecology and underlie many of the key intuitions and ideas of community ecology. CRMs consist of consumers, or species, that grow by consuming resources, while simultaneously modifying resource abundances. A defining feature of CRMs is the idea of trophic layer separation, namely that interactions between species are primarily mediated by competition for resources rather than direct interactions between species themselves. CRMs also generally ignore the effect of stochasticity, assuming that ecological selection and immigration from regional species pools are the driving force of community ecology. Despite these assumptions, recent work suggests CRMs can be used to describe a wide variety of laboratory experiments and naturally occurring empirical patterns in diverse ecosystems~\cite{marsland2020minimal,goyal2021ecology,dal2021resource, ho2022competition, ho2022resource}.
 
Over the last few years, there has been a flurry of works analyzing CRMs from the perspective of statistical physics~\cite{tikhonov2017collective,marsland2019available,advani2018statistical,cui2020effect,moran2022defining}. A key theoretical development in this regard has been the realization that diverse ecosystems, where the number of resources and species is large, can be analyzed using the cavity method (and dynamical mean field theory) \cite{bunin2017ecological, advani2018statistical, roy2019numerical, cui2021diverse, de2023many}.These works have found that the key quantities that shape collective ecosystem properties are susceptibilities that measure how species and resource abundances change in response to small environmental perturbations. \comment{This suggests that, just as in physics, characterizing the response of systems to changes in external parameters represents a powerful way of analyzing complex ecosystems. Inspired by this observation, we develop a perturbative framework for the steady state behavior of generalized consumer-resource models. Here, we limit ourselves our analysis to static perturbations and ask how steady-state properties before and after environmental perturbations are related. To do so, we derive explicit expressions for static susceptibilities, namely functions that relate the magnitude the external perturbations (e.g. a change in the carrying capacities of resources and death rates of species) to the changes in ecologically measurable variables such as species and resource abundances. } 

Our framework can be used to analyze a wide class of systems, including models with non-reciprocal interactions~\cite{blumenthal2023phase}, microbial ecosystems with cross-feeding~\cite{goldford2018emergent,goyal2018diversity}, and non-linear growth/consumption rates. Like in the cavity method, a central role is played by generalized susceptibility matrices that characterize how species and resource abundances respond to environmental perturbations. We show that these susceptibility matrices have a natural interpretation as maps between four distinct vector spaces (resource abundances, resource fluxes, species abundances, species fluxes) and define a (non-commutative) geometry that underlies CRMs. Furthermore, we show that this geometry has a natural mathematical representation in terms of a generalized ``Einstein notation'' that explicitly encodes the structure of these four vector spaces.

\comment{The resulting framework is surprisingly mathematically rich and provides intuitive geometric interpretations for understanding abstract ecological concepts such as niches, species packing, and environmental feedbacks. In particular, we use our perturbative framework to show that around any steady-state, no matter how complex the ecological dynamics, the underlying ecology can be captured by a simple MacArthur CRMs with appropriately chosen parameters. A direct implication of this mapping is that it is extremely difficult to distinguish cooperative and competitive interactions by measuring species' responses to external perturbations. The reason for this is that cooperative interactions can often be ``absorbed'' into effective parameters in competitive models, and the values of the net species interaction parameters remain competitive. Finally, we also argue that the response to external perturbations provides a natural framework for thinking about niches as ``collective modes'' composed of many resources.}

The paper is organized as follows. We start by introducing the basic ideas behind our perturbative framework in the context of Generalized MacArthur Consumer-Resource Models, including variants with non-reciprocal interactions. We then extend this construction to understand environmental perturbations in generalized consumer-resource models, including CRMs with non-linear growth and impact functions. We then illustrate these ideas on two examples inspired by microbial ecology: a model system of two microbes that interact via cross-feeding and a comparison of the effect of changing pH in specialist and generalist denitrifying bacterial communities. \comment{We conclude by discussing the broader implications of this framework for understanding ecological processes.}
\section{Analysis of MacArthur's Consumer-Resource Model}\label{sec:I}

In this section, we develop the basic formalism and geometry of our linear response framework, focusing on a generalization of MacArthur's CRM with non-reciprocal interactions.This simple setting is useful for introducing the natural mathematical objects in our construction, namely the four vector spaces in which steady states live, as well the mappings between these spaces via susceptibility matrices. It also allows us to introduce the Einstein notation used throughout the rest of the text, as well as the basic geometry of these systems. \comment{Throughout, we discuss how non-reciprocal interactions between species and resources affect the underlying geometry of these vector spaces.}

\begin{figure*}[ht!]
    \centering
    \includegraphics[width=0.9\textwidth]{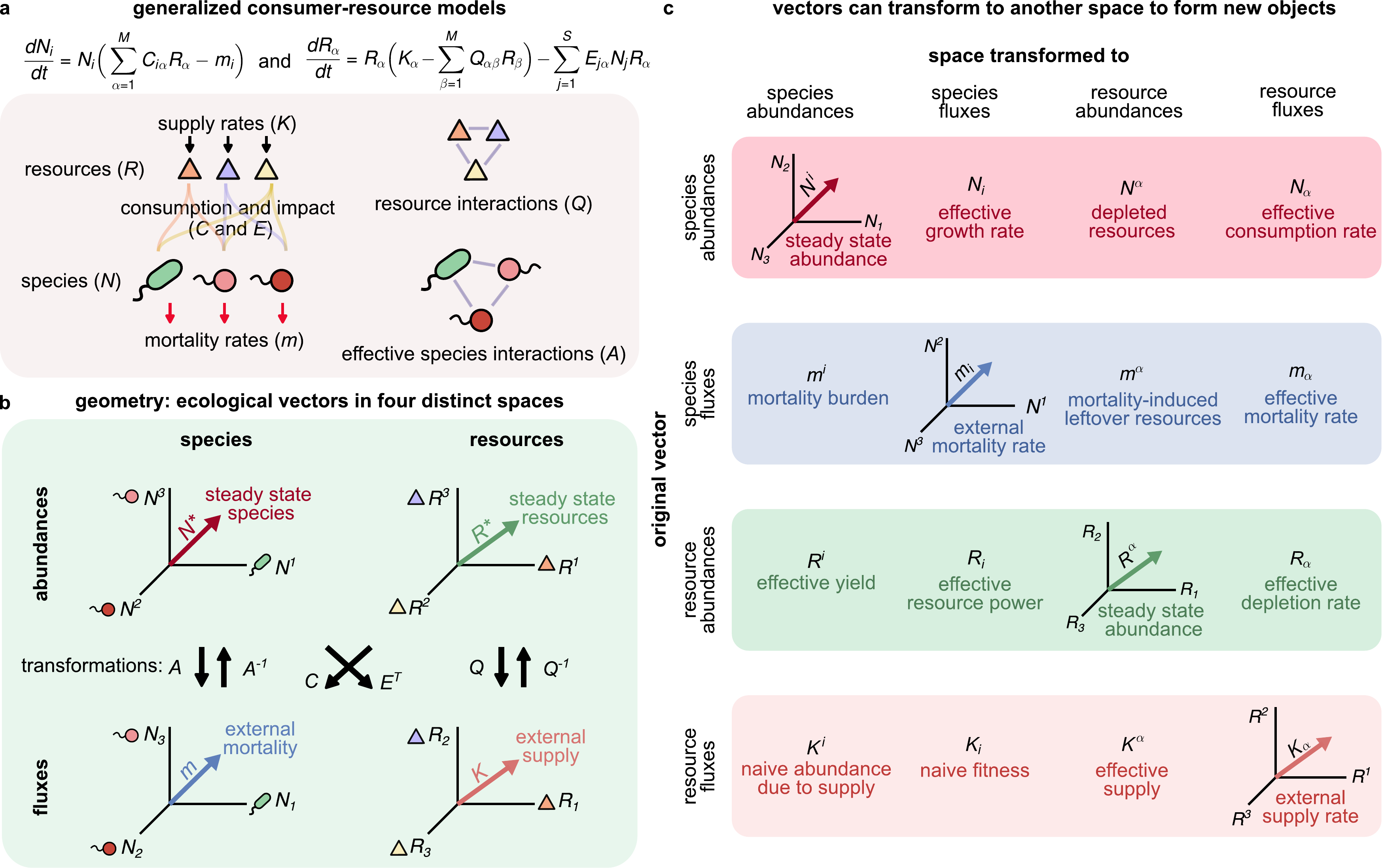}
    \caption{\footnotesize\textsf{\textbf{Ecological variables and interactions as geometric objects in four vector spaces.} 
    (a) Equation and schematic defining key quantities in consumer-resource models with non-reciprocal interactions $C \neq E$. The entries of the effective species-species interaction matrix $A_{ij}=[CQ^{-1}E^T]_{ij}$ measure how species $i$ responds to changes in species $j$. 
    (b) There are four canonical vector spaces for ecological variables and parameters:  species abundances, resource abundances, resource fluxes (quantities with units of resource abundance per unit time), and species fluxes (quantities with units of species abundance per unit time). Abundances are denoted with raised indices and fluxes with lowered indices. Vectors can be mapped between spaces using the illustrated transformations. 
    (c) Interpretation of transformed vectors. Non-native (off-diagonal) transformations reveal previously hidden, ecologically relevant quantities, including the effective supply $K^\alpha$ and mortality burden $m^i$. 
     }}      
    \label{fig:1}
\end{figure*}

\subsection{Dynamics and relation to Generalized Lotka-Volterra}
 In order to gain intuition and develop our formalism, it is helpful to start by analyzing the generalization of MacArthur's CRM shown in Fig.~\ref{fig:1}a . In this model, the ecosystem is initialized with a regional species pool of $S$ species or consumers, with abundances $N_i$,  that can utilize  $M$ distinct resources, with abundances $R_\alpha$. Species are defined by their consumer preferences, $C_{i \alpha}$, which encode how  a resource $\alpha$ contributes to the growth rate of species $i$,  impact vectors $E_{i \alpha}$, which encode how the abundance of resource $\alpha$ is affected by species $i$, and an intrinsic mortality rate $m_i$ for each species. In the absence of consumers, the resource dynamics are described by the Generalized Lotka-Volterra equations with resource-resource interaction matrix $Q_{\alpha \beta}$ and effective external supply rate  (carrying capacity) $K_\alpha$. Throughout, we assume that $Q_{\alpha \beta}$ is full rank and invertible. These dynamics take the form of the coupled ordinary differential equations
\begin{equation}
\begin{aligned}
\dv{N_i}{t} &= N_i \qty(\sum_\alpha C_{i \alpha} R_\alpha -m_i) \\
\dv{R_\alpha}{t} &= R_\alpha \qty(K_\alpha - \sum_\beta Q_{\alpha \beta} R_\beta) - \sum_j E_{j \alpha} N_j R_{\alpha}.
\end{aligned}\label{Eq:GMCRM}
\end{equation}
For the special case when $Q_{\alpha \beta}$ is the identity matrix and ${C_{i \alpha}=E_{i \alpha}}$, this reduces to the usual MacArthur Consumer-Resource Model that has been extensively studied in the ecological literature~\cite{chesson1990macarthur}.

To gain some intuition about these equations, it is helpful to consider what happens when the resource dynamics reach steady state. In this case, the species abundances \comment{fully depend on} the resources. Setting ${\dv{R_\alpha}{t}=0}$, and substituting into the first equation above gives the effective Generalized Lotka-Volterra  equations for the consumer species (see Appendix~\ref{Appendix:GLVderivation}):
\be
\dv{N_i}{t} = N_i \qty(\mathcal{K}_i - \sum_j A_{ij} N_j),
\label{Eq:GLV}
\ee
where we have defined the effective species-species interaction matrix 
\begin{align}
A_{ij}&= \sum_{\alpha\beta} C_{i \alpha} Q^{-1}_{\alpha \beta} E_{j \beta}
\label{Eq:GLV-A}
\end{align}
and written the effective carrying capacity of a species as the difference
\begin{align}
\mathcal{K}_i &=\sum_{\alpha, \beta} C_{i \alpha} Q^{-1}_{\alpha \beta}  K_\beta - m_i
\label{Eq:GLV-K}
\end{align}
between the ``naive'' fitness (first term) and the mortality rate $m_i$.
 In the expressions above, all sums over resources are restricted to non-extinct resources (Fig.~\ref{fig:1}a). 

When ${C_{i \alpha}=E_{i \alpha}}$ and  $Q_{\alpha \beta} $ is the identity, this model reduces to MacArthur's original model and the effective interaction matrix $A_{ij}$ defined in Eq.~\eqref{Eq:GLV-A} is symmetric, indicating that species interact reciprocally. In other words, how species $i$ influences species $j$ is identical to how species $j$ influences species $i$. The system is also guaranteed to reach a unique steady state that corresponds to a minimum of a constrained optimization problem~\cite{mehta2019constrained, marsland2020minimum}. For this reason, the MacArthur CRM shares many properties with equilibrium systems. 

In contrast, when the consumption matrix $C_{i \alpha}$ is different from the impact matrix $E_{i \alpha}$ , the resulting species-species interaction matrix in Eq.~\eqref{Eq:GLV-A} is no longer symmetric (i.e., $A_{ij} \neq A_{ji}$). This indicates that species no longer interact reciprocally, a hallmark of nonequilibrium physics~\cite{fruchart2021non,ivlev2015statistical}. As a result, the dynamics no longer have any simple description in terms of optimization and can even become chaotic when $C_{i \alpha}$ and $E_{i \alpha}$ are sufficiently different~\cite{blumenthal2023phase}. 

\subsection{Steady states} 
In what follows, we focus on steady-state solutions of generalized CRMs of the form described in Eq.~\eqref{Eq:GMCRM}. We assume that these ecosystems reach a steady-state fixed point (i.e., are not chaotic or periodic), and wish to characterize how the steady-state resource and species abundances depend on environmental parameters such as the resource supply rates $K_\alpha$ and species mortality rates $m_i$. In this section, we derive simple equations for this dependence, which in turn motivate the more general framework presented in the next section and the rest of the paper.

Let us denote the number of surviving species and resources at the steady state we wish to characterize as $S^*$ and $M^*$, respectively. It follows from Eq.~\eqref{Eq:GMCRM} that the steady-state abundances of surviving species, $N_j^*$, and surviving resources, $R_\alpha^*$, satisfy the $M^*+S^*$ linear equations
\begin{equation}
\begin{aligned}
\sum_{\beta=1}^{M^*} C_{i \beta} R_\b^*&= m_{i}\\
\sum_{\beta=1}^{M^*} Q_{\alpha \beta}R_\beta^*  +\sum_{j=1}^{S^*} E_{j \alpha} N_j^* &= K_\alpha,
\end{aligned}\label{Eq:GMCRM-SS}
\end{equation}
where the resource index $\alpha$ and the species index $i$ index run over the $M^*$ and $S^*$ surviving resources and species, respectively. 

Since we wish to characterize how the \emph{steady-state abundances} of \emph{surviving species and resources} depend on environmental perturbations, for notational simplicity we will adopt the convention of dropping asterisks. Unless explicitly noted, all steady-state equations should be assumed to exclusively involve the $M^*$ resources and $S^*$ species that survive at the fixed point. With this notational change, we can rewrite Eq.~\eqref{Eq:GMCRM-SS} in matrix form as
\be
\mqty[
0 & C\\
E^T & Q
]
\mqty[
 \vec{N}\\
 \vec{R}
]
=
\mqty[
 \vec{m}\\
 \vec{K}
],
\ee
where $C$ and $E$ are $S^* \times M^*$ dimensional growth and impact matrices (i.e., matrices restricted to surviving species and resources), $Q$ is the the $M^* \times M^*$ dimensional resource interaction matrix, $\vec{R}$ and $\vec{K}$ are $M^*$ dimensional column vectors of resource abundances and supply rates, respectively, and $\vec{N}$ and $\vec{m}$ are $S^*$ dimensional vectors of species abundances and mortality rates, respectively.

\begin{figure*}[t!]
    \centering
    \includegraphics[width=0.9\textwidth]{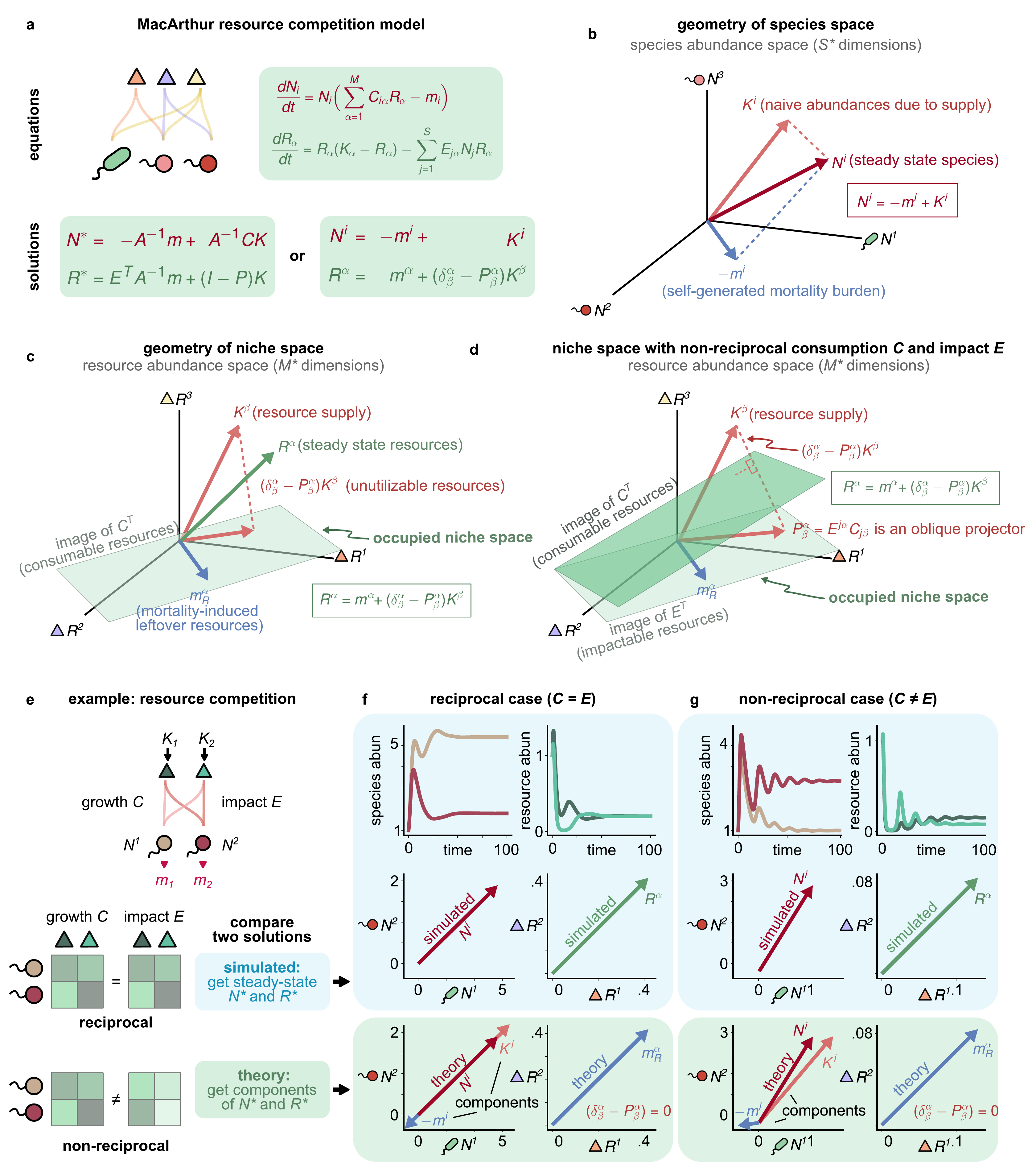}
    \caption{\footnotesize\textsf{\textbf{Geometry of niche space.}
    Geometric decomposition of the steady-state resource abundance vector for the Generalized MacArthur Consumer-Resource Model (with $Q=I$).  (a) Equations for dynamics and the steady-state abundances of surviving resource and species. (b) The vector of steady-state species abundances $N^i$ is the difference of naive abundance due to supply $K^i$ and the self-generated mortality burden $m^i$. (c) Geometry of steady-state resource abundances in models with (c) reciprocal interactions ($C=E$) and (d) and non-reciprocal interactions $(C \neq E)$. The steady-state resource abundance vector $R^\alpha$ can be decomposed into a vector sum of unutilizable resources, $(I-P_\beta^\alpha )K^\beta$, and mortality-induced leftover resources, $m_R^\alpha$. (e) Comparison between numerics and predictions of geometric framework for a 2 species, 2 resource MacArthur Consumer-Resource Model model with reciprocal ($C=E$) and non-reciprocal ($C\neq E$) interactions. (see Appendix ~\ref{Appendix:Methods} for details).
     }}
    \label{fig:2}
\end{figure*}

As shown in Appendix~\ref{Appendix:SSExpression}, a straightforward application of the block-inversion formula for $2 \times 2$ matrices allows us to solve this equation. In particular, 
one can show that 
\begin{align}
\mqty[
0 & C\\
E^T & Q
]^{-1}
= 
\mqty[
-I  & A^{-1} C\\
Q^{-1}E^T & I- P
]\mqty[
A^{-1} & 0\\
0& Q^{-1}
],
\label{Eq:blockmatrixinverse}
\end{align}
where $A$ is the $S^* \times S^*$ dimensional species-species interaction matrix for surviving species [defined in Eq.~\eqref{Eq:GLV-A} but now restricted to surviving species and resources] and $P$ is an $M^* \times M^*$ dimensional ``projection'' matrix
\be
P=Q^{-1}E^TA^{-1}C.
\label{Eq:Projectionmatrix-block}
\ee
\comment{Using Eq.~\eqref{Eq:GLV-A}, it is easy to show that $P$ satisfied the projection equation $P^2=P$. A projector
is a linear operator that takes a vector and projects it onto a linear subspace (see Fig.~\ref{fig:2}.). When species interactions are non-reciprocal ($C \neq E$ or  $Q$ is non-symmetric), $P \neq P^T$ indicating that $P$ is an oblique projector. In contrast, for the MacArthur CRM ($C=E$ and $Q$ symmetric), $P=P^T$ and $P$ is an orthogonal projector.}

We can now use the expression for the inverses to write expressions for the steady-state species and resource abundances.
 Using Eq.~\eqref{Eq:GMCRM-SS}, we find that
 
\begin{equation}
\begin{aligned}
\vec{N}& =  A^{-1}[CQ^{-1} \vec{K} -\vec{m}] = A^{-1}\vec{\mathcal{K}} \\
\vec{R}& =  Q^{-1}E^T A^{-1} \vec{m} + (I-P) Q^{-1}\vec{K}.
\end{aligned}\label{Eq:GMCRM-SS-final}
\end{equation}

Comparing the first equation to Eqs.~\eqref{Eq:GLV} and~\eqref{Eq:GLV-K}, we recognize it as just the steady-state condition for the effective Lotka-Volterra equation, where the effective species carrying capacity $\vec{\mathcal{K}}$ captures both the naive fitness and the effect of intrinsic mortality $\vec{m}$ typically ignored in the Generalized Lotka-Volterra model. Furthermore, we see that in the second equation, the quantity $PQ^{-1}\vec{K}$ is the projection of the resource supply vector onto a subspace defined by the consumption and impact vectors of surviving species contained in $P$. 

These  are our first hints that a more general mathematical structure is encoded in the elements of this simple equation. In Appendix~\ref{Appendix:MCRM-susc}, we show that the matrix inverse appearing in Eq.~\eqref{Eq:blockmatrixinverse} also has a natural interpretation as a block susceptibility matrix that encodes how the steady-state resource abundances $\vec{R}$ and species abundances $\vec{N}$ change in response to small perturbations in $\vec{K}$ and $\vec{m}$. We will use this observation when generalizing our results beyond the simple models considered in Eq.~\eqref{Eq:GMCRM}.

\subsection{Defining vector spaces}
The formulae derived in the previous section link four distinct types of quantities that appear in the dynamical equations [see Eq.~\eqref{Eq:GMCRM} and Fig.~\ref{fig:1}b]: \\
\begin{enumerate}
\item Quantities such as $R_\alpha$ that measure resource abundances
\item Quantities such as $K_\alpha$ that measure fluxes per unit resource (i.e., quantities labeled by resources that have units of inverse time)
\item Quantities such as $N_i$ that measure species abundances
\item Quantities such as $m_i$ that measure fluxes per unit species (i.e., quantities that are labeled by species and have units of inverse time). 
\end{enumerate}
Each of these four types of quantities can be thought to describe one of four distinct vector spaces (Fig.~\ref{fig:1}b). Furthermore, closer inspection of Eq.~\eqref{Eq:GMCRM-SS-final} shows that each of these quantities can be mapped between spaces using a combination of the matrices $C$, $E$, $Q$, $Q^{-1}$, $A$, and $A^{-1}$ (Fig.~\ref{fig:1}b). For example,  the first equation in Eq.~\eqref{Eq:GMCRM-SS-final} implies that the naive fitness $CQ^{-1} \vec{K}$ has the same units as $\vec{m}$, suggesting that $CQ^{-1}$ maps vectors like $\vec{K}$ that live in resource flux space to vectors like $\vec{m}$ in species flux space. Furthermore, since both sides of this equation must have the same units, it also implies that one can map back and forth between vectors living in species flux and species abundance space using $A$ and $A^{-1}$, respectively. Similarly, the second equation in Eq.~\eqref{Eq:GMCRM-SS-final} implies that $E^TA^{-1}$ maps vectors like $\vec{m}$ that live in species flux space to vectors like $\vec{K}$ in resource flux space. It also implies that one can map between the resource flux  and resource abundance spaces using $Q$ and $Q^{-1}$. 

In Fig.~\ref{fig:1}b, we graphically illustrate how different matrices can be combined to map vectors between these four spaces. As an example, using Fig.~\ref{fig:1}b, we see that the projector $P=Q^{-1}E^TA^{-1}C$, can be naturally decomposed into a series of linear transformations that cycle from resource abundance space back to it, taking the route: resource abundance $\rightarrow$ species flux $\rightarrow$ species abundance $\rightarrow$ resource flux $\rightarrow$ resource abundance. We note that similar mappings have been found in the context of generalized susceptibilities in metamaterials~\cite{rocks2022integrating} and overparameterized regression~\cite{rocks2021geometry}. \comment{Below, we discuss how this mathematical structure allows us 
to give intuitive geometric definitions of abstract ecological concepts such as niches, species packing, and environmental feedbacks.}



\subsection{Generalized Einstein notation}

Thus far, the mathematical notation we have employed does not distinguish between the four natural spaces that characterize the ecological steady states of Generalized MacArthur CRMs. Nor does it reflect the structure of the transformations between these spaces (Fig.~\ref{fig:1}b) . To rectify this, we introduce generalized Einstein notation that better captures the underlying geometry. This new notation allows us to easily distinguish between spaces and can also be used to quickly perform calculations without the need to use block matrix inversion formulas (see Appendix~\ref{Appendix:MCRM-EinNot}).

A key feature of our notation will be the introduction of upper and lower indices (Fig.~\ref{fig:1}c). In particular, we use: 
\begin{enumerate}
\item Upper Greek indices for quantities with units of resource abundance (e.g., $R^\alpha$)
\item Lower Greek indices for quantities with units of resource flux (e.g., $K_\alpha$ )
\item Upper Latin indices for quantities with units of species abundance (e.g., $N^i$)
\item Lower Latin indices for quantities with units of species flux (e.g.,  $m_i$).
\end{enumerate}
As in conventional Einstein notation, will also assume that repeated upper and lower indices of the same kind are summed over (surviving species or resources).

A key feature of the underlying mathematical structure illustrated in Fig.~\ref{fig:1}b is that we can map vectors between these four spaces using the matrices $Q$, $A$, $C$, and $E$ [see also Eqs.~\eqref{Eq:GMCRM-SS} and~\eqref{Eq:GMCRM-SS-final}]. All four of these matrices map abundances (upper indices) to fluxes (lower indices). For this reason, they must have two lower indices and take the form $Q_{\alpha \beta}$, $E_{i \alpha}$, $C_{i \beta}$, and $A_{ij}$. In contrast, the two matrices $Q^{-1}$ and $A^{-1}$ both map fluxes to abundances. For this reason they must have two upper indices. In index notation, $Q^{-1}$ becomes $Q^{\alpha \beta}$ and $A^{-1}$ becomes $A^{ij}$.  In Fig.~\ref{fig:1}c, we graphically illustrate each of the sixteen key ecological quantities as vectors that live in these four spaces, along with a brief ecological interpretation of each vector. We will use these 16 vectors to provide a geometric interpretation of CRMs and identify emergent ecological quantities that underlie ecosystem steady states. A detailed discussion of these vectors and mappings is provided in Appendix \ref{Appendix:EinsteinOperators}.

\section{Geometry of Niche and Species Space} 
\label{sec:geom-int-MCRMs}
\subsection{Derivation of geometry}
A powerful consequence of being able to map quantities between the four vectors spaces outlined in the last section is that it allows us to develop a simple geometric picture for understanding ecological steady states  (Figs.~\ref{fig:2}a-d). Our starting point is Eq.~\eqref{Eq:GMCRM-SS-final} for the steady-state resource and species abundances, which we can rewrite in Einstein notation as
\begin{equation}
\begin{aligned}
N^i &=A^{ij}C_{i \alpha}Q^{\alpha \beta}K_\beta -A^{ij} m_j \\
R^\alpha& =Q^{\alpha \beta} E_{j \alpha} A^{jk}m_k + (\delta^\alpha_\beta -P^\alpha_\beta) Q^{\beta\gamma}K_\gamma.
\end{aligned}
\end{equation}
In terms of vectors introduced in Fig.~\ref{fig:1}c and Eqs.~\eqref{Eq:4Rvec}-\eqref{Eq:4mvec},
these equations takes the compact form
\begin{equation}
\begin{aligned}
N^i &=K^i -m^i\\
R^\alpha& =m^\alpha + (\delta^\alpha_\beta -P^\alpha_\beta) K^\beta.
\end{aligned}\label{Eq:GMCRM-ess}
\end{equation}

Both of these equations have a beautiful ecological interpretation. The first equation states that the steady-state species abundances $N^i$ is the difference between two vectors: the ``naive abundance due to supply'' $K^i$ and the ``mortality burden'' $m^i$. By inspection, we see that $K^i$ represents the abundances that each species would reach when the only limitation to species growth is the amount of supplied resources. However, the actual abundance is reduced by the mortality burden $m^i$, which captures the decrease in species abundances due to non-zero mortality rates (i.e., $m_i \neq 0$ for any species).
Note that $K^i$ and $m^i$ are emergent ecological quantities that do not directly occur in our original model. Geometrically, this equation can be visualized in species space as shown in Fig.~\ref{fig:2}b. 

The second equation in Eq.~\eqref{Eq:GMCRM-ess} states that the steady-state resource abundance $R^\alpha$ also has two components: the ``mortality-induced leftover resources'' $m^\alpha$, and the ``unutilizable resources'',  ${(\delta^\alpha_\beta -P^\alpha_\beta) K^\beta}$ (see  Fig.~\ref{fig:2}c-d). To develop an intuition for these vectors, it is helpful to draw on some of the ecological intuition developed above. 

We first justify the interpretation of $(\delta^\alpha_\beta -P^\alpha_\beta) K^\beta$ as resources that cannot be utilized by species currently living in the ecosystem. Our starting point is the observation that $K^\beta$ is simply the ``effective supply'' of resources, i.e., the steady-state resource abundances in the absence of any consumers. This can be seen by setting either the consumption or impact vectors to zero, $C_{i \alpha}=0$ or $E_{i \alpha}=0$, in which case Eq.~\eqref{Eq:GMCRM-ess} simplifies to ${R^\alpha = K^\beta=Q^{\alpha \beta} K_\alpha}$.

Next,  we observe that the projector ${P^\alpha_\beta = E^i_\beta C_i^\alpha}$ acts on resource vectors in two steps. First, it projects resource vectors onto the subspace of `consumable resources'' defined by the row space of $C$ (or the image of $C^T$).
Next, $P^\alpha_\beta$ projects the result onto the subspace of ``impactable resources'' defined by the row space of $E$ (or the image of $E^T$). 
The interaction of these two steps defines the space of ``utilizable resources,'' (which is a strict subspace of the impactable resources). 
In other words, resouces may only be utilized by the current community of species in the ecosystem if they are both consumable by the species \textit{and} this consumption impacts the abundance of the resources.
For the case where interactions are reciprocal, these two spaces coincide, and $P_\alpha^\beta$ is an orthogonal projector (Fig.~\ref{fig:2}c). For non-reciprocal interactions, these spaces are distinct and $P_\alpha^\beta$ is an oblique projector (Fig.~\ref{fig:2}d), first orthogonally projecting resource vectors onto the row space of $C$ before projecting orthogonally off of this space onto the row space of $E$.
Since $(\delta_\alpha^\gamma -P_\alpha^\gamma)P_\gamma^\beta=0$, this implies that ${\delta_\alpha^\gamma -P_\alpha^\gamma}$ is a projector onto the space orthogonal to the space of utilizable resoruces, i.e, the space of unutilizable resources, justifying our naming convention. 

Next,  we turn to the ``mortality-induced leftover resources'' $m^\alpha$. First, note that the vector $m^\alpha$ lives in the image of $P_\alpha^\beta$, namely 
\begin{align}
P^\alpha_\beta m^\beta &=m^\alpha.
\end{align}
For this reason, the vector $m^\alpha$ is composed only of utilizable resources (Figs.~\ref{fig:2}c-d). This allows us to interpret $m^\alpha$ as the contribution to the steady-state resource abundance of resources that can be potentially utlized by the community but are left unconsumed due to species mortality. This shows that the unutilized niches in the ecosystem -- realizable but unoccupied -- can be encoded in a single emergent vector quantity, $m^\alpha$. Thus, the steady-state resource abundance vector $R^\alpha$ naturally decomposes into a mortality-induced niche vector $m^\alpha$ and an unutilizable niche vector $(\delta^\alpha_\beta -P^\alpha_\beta) K^\beta$ (Figs.~\ref{fig:2}c-d).

\subsection{Example: resource competition with reciprocal and non-reciprocal interactions}

To see the geometric picture in action, it is useful to consider a concrete example. For ease of visualization, we focus on ecosystems consisting of 2 resources and 2 species whose dynamics are described by Eq.~\eqref{Eq:GMCRM} (see Figs.~\ref{fig:2}e-g). We compare and contrast two different communities, one where species interact reciprocally ($C=E$, Fig.~\ref{fig:2}f) and another where species interactions are non-reciprocal ($C \neq E$, Fig.~\ref{fig:2}g). In both communities, resources are supplied at rates $K_1$ and $K_2$ and species have mortality rates $m_1$ and $m_2$ (see Appendix ~\ref{Appendix:Methods} for details). Both ecosystems also have identical consumer preference matrices $C$. The only difference between the ecosystems are the impact vectors $E$ (Fig.~\ref{fig:2}e, bottom). 

To analyze these communities, we ran simulations where we numerically solved for the steady-state species and resource abundance vectors $N^i$ and $R^i$, evolving the dynamics from arbitrarily chosen initial conditions (Figs.~\ref{fig:2}f-g, blue box), and compared the steady-state values to predictions derived from the analytic expressions in Eq.~\eqref{Eq:GMCRM-ess} (Fig.~\ref{fig:2}, green box). As expected, the two results are in good agreement. Our geometric picture also yields a natural decomposition $N^i = K^i + m^i$ of the steady-state abundance into the naive abundances , $K^i$ [Eq.~\eqref{Eq:4Kvec}, Fig.~\ref{fig:1}], and the mortality burden $m^i$ [Eq.~\eqref{Eq:4mvec}, Fig.~\ref{fig:1}'.  As emphasized above, both of these vectors are ``emergent quantities'' that depend strongly on species consumer preferences and impact vectors. The naive abundances $K^i$ account for the fact that the effective supply rate is shaped both by the amount of supplied resources and resource consumption. The mortality burden $m^i$ incorporates not only the naive mortality rate, but also death due to competition between species.

Notice that the geometric relationship between $K^i$ and $m^i$ is qualitatively different depending on whether species interact in a non-reciprocal manner (Figs.~\ref{fig:2}e-g). For the community with reciprocal interactions, $K^i$ and $m^i$ are collinear whereas for the community with non-reciprocal interactions, these are at an obtuse angle to each other. This is the geometric manifestation of the fact that  species that interact non-reciprocally  decouple growth rates from consumption.  As we will see in future examples, such obtuse angles turn out to be a generic feature of non-reciprocity (see Appendix~\ref{Appendix:Geom-Angles}).  Finally, we note that for both communities ${(\delta^\alpha_\beta -P^\alpha_\beta) K^\beta=0}$ because all the resources are utilizable. As a result, the steady-state resource abundance is given entirely by the mortality-induced leftoever resources $m^\alpha$ (see  Figs.~\ref{fig:2}f-g).

\subsection{Lessons from MacArthur Models} 
Combining the results from Sec.~\ref{sec:I}, we arrive at the following lessons about steady states in MacArthur's consumer-resource models and how they can be interpreted geometrically:
\begin{itemize}
\item There are four distinct kinds of steady-state quantities: species and resource abundances and fluxes (Fig.~\ref{fig:1}a). These quantities describe four distinct vector spaces (Fig.~\ref{fig:1}b).
\item Vectors in each space can mapped to any other space through ecological susceptibilities -- consumption $C$, impact $E$, species interactions $A$ and $A^{-1}$, and resource interactions $Q$ and $Q^{-1}$ -- decomposed as in Fig.~\ref{fig:1}b. Each of these 12 mapped vectors is an emergent ecological quantity with a clear interpretation (Fig.~\ref{fig:1}c).
\item Ecological steady states can be decomposed in terms of 4 of the 12 emergent vector quantities that summarize the effects of ecological competition, niches, and mortality, as shown in Eq.~\eqref{Eq:GMCRM-ess} and Figs.~\ref{fig:2}b-d.
\item Relations between ecological quantities and susceptibilities can be condensed using generalized Einstein notation [Eqs.~\eqref{Eq:4Rvec}-\eqref{Eq:GMCRM-ess}], further clarifying the geometry underlying these relations. 
\end{itemize}

\begin{figure*}[t!]
    \centering
   \includegraphics[width=0.9\textwidth]{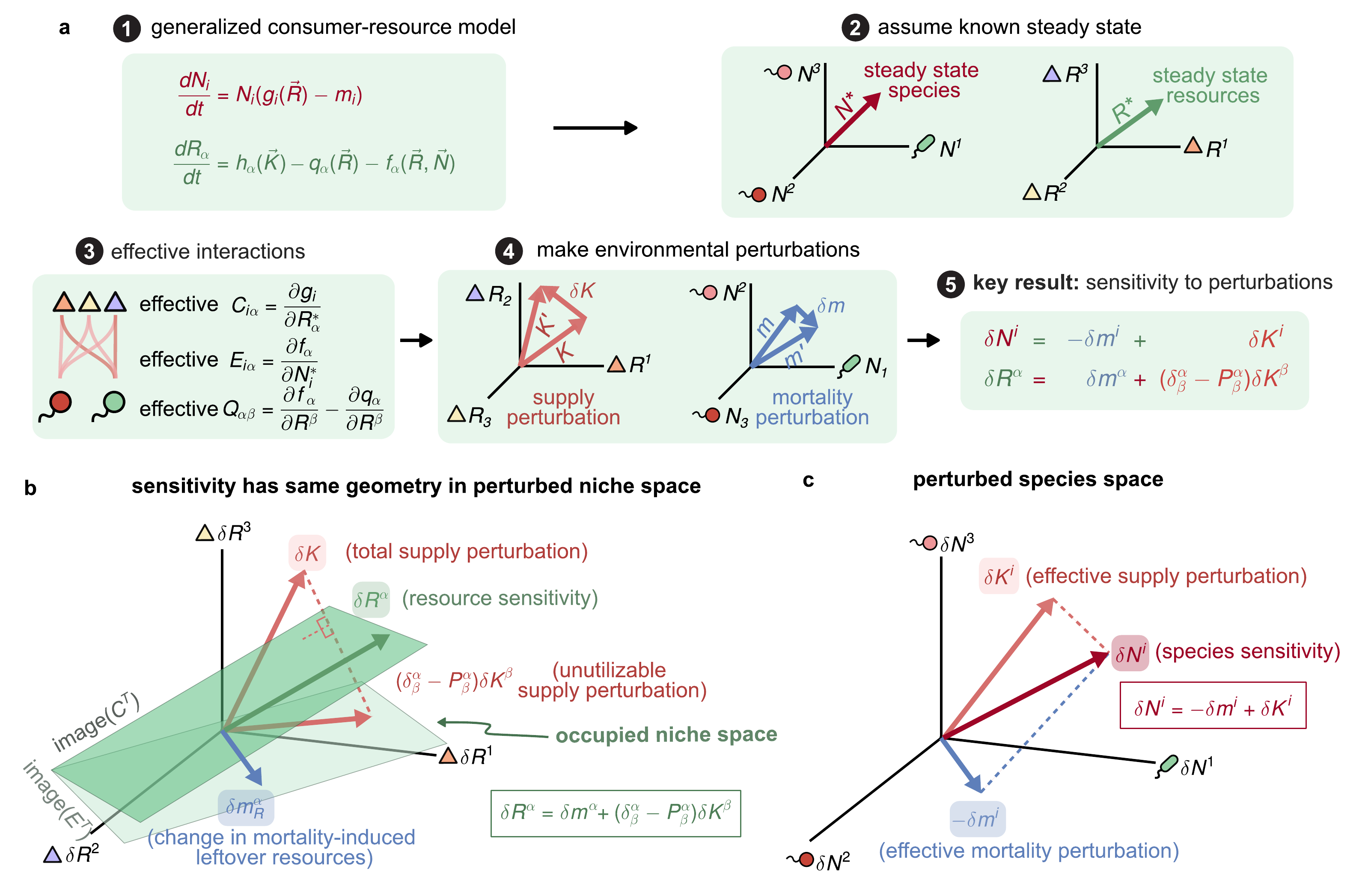}
    \caption{\footnotesize\textsf{\textbf{Universal geometry predicts sensitivity to environmental perturbations.} 
    (a) Near the steady state of any generalized consumer-resource model, environmental perturbations in supply $\vec{\delta K}=\vec{K'}-\vec{K}$ (pink) and mortality $\vec{\delta m}=\vec{m'}-\vec{m}$ (blue) result in changes in the steady-state resource and species abundances, $\delta R^\alpha$ and $\delta N^i$ via effective interactions $C$, $Q$ $E$, $A$ and projector $P$.
    (b-c) To linear order, the steady-state responses $\vec{\delta R}$ and $\vec{\delta N}$ have a geometrical decomposition in terms of the perturbations $\delta \vec{K}$ (pink) and $\delta \vec{m}$ (blue) analogous to the decomposition for steady states in the MacArthur Consumer-Resource Model shown in Fig.~\ref{fig:2}.
    }}
    \label{fig:3}
\end{figure*}

\section{Characterizing responses to environmental perturbations in generalized Consumer-Resource Models}
Thus far, we have restricted our analysis to MacArthur's Consumer-Resource Model (and its generalization to non-reciprocal interactions). This simple setting allowed us to introduce our geometric framework and the steady-state ecological quantities of interest. In this section, we extend our analysis to understanding environmental perturbations in generalized CRMs.  


\subsection{Generalized Consumer-Resource Models}
\label{sec:GCRM-intro}
Generalized CRMs describe ecosystems containing $S$ species with abundances $N_i$, and $M$ resources with abundances $R_\a$, whose dynamics takes the form 
\begin{equation}
\begin{aligned}
\dv{N_i}{t} &= N_i\qty(g_i(\vec{R}) - m_i) \\
\dv{R_\alpha}{t} &= h_\alpha(\vec{K}) - q_\alpha(\vec{R}) - f_{\alpha}(\vec{R}, \vec{N}),
\end{aligned}\label{Eq:generalRD}
\end{equation}
where $g_i(\vec{R})$ is a function which describes how the growth rate of species $i$ depends on resource abundances, $h_\alpha(\vec{K})$ is the rate at which resource $\a$ is supplied to the ecosystems, $q_\alpha(\vec{R})$ encodes the resource dynamics in the absence of species, and $f_\a(\vec{R},\vec{N})$  describes the rate at which resource $\a$ is produced or consumed by species in the ecosystems.  Like in all CRMs, competition between species is mediated by the resources rather than through direct interactions. Mathematically, this is just the statement that the growth rate functions $g_i(\vec{R})$ are only a function of the resource abundances and are independent of species abundances. This strict trophic layer structure is a hallmark of all CRMs.

We restrict ours analysis to ecosystem where the dynamics reach a steady state, $\vec{R}^*$ and $\vec{N}^*$. We are interested in characterizing the response of such a system to environmental perturbations. That is, if we slightly change
the resource supply rates ${K_\a \rightarrow   K_\a + \delta \tilde{K}_\a}$ and the species mortality ${m_i \rightarrow m_i +  \delta m_i}$, how will the resulting steady states change, i.e.,  ${{R^\alpha}^* \rightarrow {R^\alpha}^* + {\delta R^\alpha}}$ and ${{N^i}^* \rightarrow {N^i}^* + {\delta N^i}}$? As in MacArthur's CRM, we will focus on surviving species and resources, dropping the asterisk $*$ indicating the survivors from our notation for convenience.

The new steady state after perturbations must also satisfy the steady-state equations,
\begin{equation}
\begin{aligned}
0&=g_i(\vec{R} + \vec{\delta R}) - m_i -\delta m_i   \\
0 &= h_\alpha(\vec{K}+\vec{\delta \tilde{K}}) - q_\alpha(\vec{R}+ \vec{\delta R}) - f_{\alpha}(\vec{R}+\vec{\delta R}, \vec{N}+ \vec{\delta N}).
\end{aligned}
\end{equation}
To calculate the linear response, we can expand these equations to first order (noting that the zeroth-order terms are zero since we are expanding about a steady state) to get
\begin{equation}
\begin{aligned}
C_{i \a} \d R^\a &= \d m_{i} \\
Q_{\a \b} \d R^\b  + E_{i \a} \d N^i &= \d K_\a,
\end{aligned}\label{Eq:GCRM-lin}
\end{equation}
where 
\begin{equation}
\begin{aligned}
C_{i\a} &= \pdv{g_i}{R^\alpha}\eval_{\vec{R}^{*}}\\
E_{i\a} &= \pdv{f_\a}{N^i}\eval_{\vec{R}^*, \vec{N}^{*}} \\ 
Q_{\a\b} &= \pdv{f_\a}{R^\b}\eval_{\vec{R}^*, \vec{N}^{*}}+ \pdv{q_\a}{R^\b}\eval_{\vec{R}^*},
\end{aligned}\label{Eq:generalCEQ}
\end{equation}
are the effective consumer preference matrix, effective impact vector matrix, and effective resource-resource interaction matrix, respectively and
\be
\d K_\a= \sum_\b \pdv{h_\a}{K^\b}\eval_{\vec{K}} \delta \tilde{K}_\b
\ee
is the effective supply perturbation vector (note that through an abuse of notation we denote the quantity on the left hand side).

These equations are essentially identical to the steady-state Eqs.~\eqref{Eq:GMCRM-SS} for MacArthur's CRM, except that they characterize environmental perturbations ($\d K_\a$ and $\d m_i$) and ecosystem responses ($\d R^\a$ and $\d N^i$) rather than the steady states themselves. We can rewrite them in matrix form to obtain
\be
\mqty[
0 & C\\
E^T & Q
]
\mqty[
 \vec{\d N}\\
 \vec{\d R}
]
=
\mqty[
 \vec{\d m}\\
 \vec{\d K}
],
\ee
and then use standard block-inversion formulae to obtain a solution for the ecosystem's steady-state response to environmental perturbations (full scheme illustrated in Fig.~\ref{fig:3}a),
\begin{equation}
\begin{aligned}
\vec{\d N}& = A^{-1} [CQ^{-1} \vec{\d K} -\vec{\d m}] \\
\vec{\d R}& =  Q^{-1} E^T A^{-1} \vec{\d m} + (1-P) Q^{-1}\vec{\d K},
\end{aligned}\label{Eq:GCRM-SS-final}
\end{equation}
where the species-species interactino matrix $A=CQ^{-1}E_T$ the generalized projector $P=Q^{-1}E^TA^{-1}C$ is defined using the effective interactions in Eq.~\eqref{Eq:generalCEQ}. Continuing the analogy with MacArthur CRMs [Eq.~\eqref{Eq:GMCRM-ess}], these solutions can be written more compactly in Einstein notation as 
\begin{equation}
\begin{aligned}
\d N^i &=\d K^i - \d m^i \\
\d R^\alpha& =\d m^\alpha + (\delta^\alpha_\beta -P^\alpha_\beta) \d K^\beta.
\end{aligned}\label{Eq:GCRM-ess}
\end{equation}
As expected, the solutions for the linear response are mathematically identical to those for MacArthur's CRM in Eq.~\eqref{Eq:GMCRM-ess}, but with the emergent effective perturbation vectors $\d K^i$, $\d K^\alpha$,  $\d m^\alpha$ and $\d m^i$ playing an analogous role to the vectors $K^i$, $K^\alpha$,  $m^\alpha$ and $m^i$, respectively.

\subsection{Geometry of linear response to environmental perturbations}
The mathematical equivalence between Eqs.~\eqref{Eq:GCRM-ess} and~\eqref{Eq:GMCRM-ess} implies that the geometric interpretation of Eq.~\eqref{Eq:GMCRM-ess} developed in Sec.~\ref{sec:geom-int-MCRMs} also applies to generalized CRMs. \comment{However, while the equations are mathematically identical, there is a conceptual difference in interpretation. In MacArthur CRMs, the geometry describes the actual abundances as a function of the supply and mortality. In contrast, in generalized CRMs, the geometry now describes changes in abundances as a function of perturbations to supply and mortality.} Geometrically, these equations can be visualized in perturbed niche and species spaces as shown in Figs.~\ref{fig:3}c-d. 

As in MacArthur's CRM, the change in steady-state resource abundances $\d R^\a$ is the sum of two components: the ``unutilizable supply perturbation'' ${(\d^\a_\b - P^\a_\b)\d K^\b}$ (Fig.~\ref{fig:3}c, dashed pink) and the change in ``mortality-induced leftoever resources'' $\d m^\a$ (blue). The first component describes changes to the effective resource supply $\d K_\a$ in resources that cannot be utilized by species in the ecosystem, whether due to an inability to consume or impact. The second describes how the abundances of utilizable resources is affected by changes in species mortality $\d m^i$.  The change in steady-state species abundance $\d N^i$ also decomposes into two quantities:  the ``effective supply perturbation'' $\d K^i$, and the ``effective mortality perturbation'' $\d m^i$ (Fig.~\ref{fig:3}d, pink and blue respectively). The first of these describes how species abundances change in the absence of changes to the mortality rate ($\d m_i=0$), while the second describes the additional burden on species due to changes in mortality. As before, the components $\d K^i$, $\d K^\a$, $\d m^i$ and $\d m^a$ are emergent quantities that utilize the effective consumer and impact vectors defined in Eq.~\eqref{Eq:generalCEQ} to propagate the effect of the actual perturbations in resource supply $\d K_\a$ and species mortality $\d m^\a$. 

\section{Examples of geometry for complex models}

\begin{figure*}[ht!]
    \centering
    \includegraphics[width=0.7\textwidth]{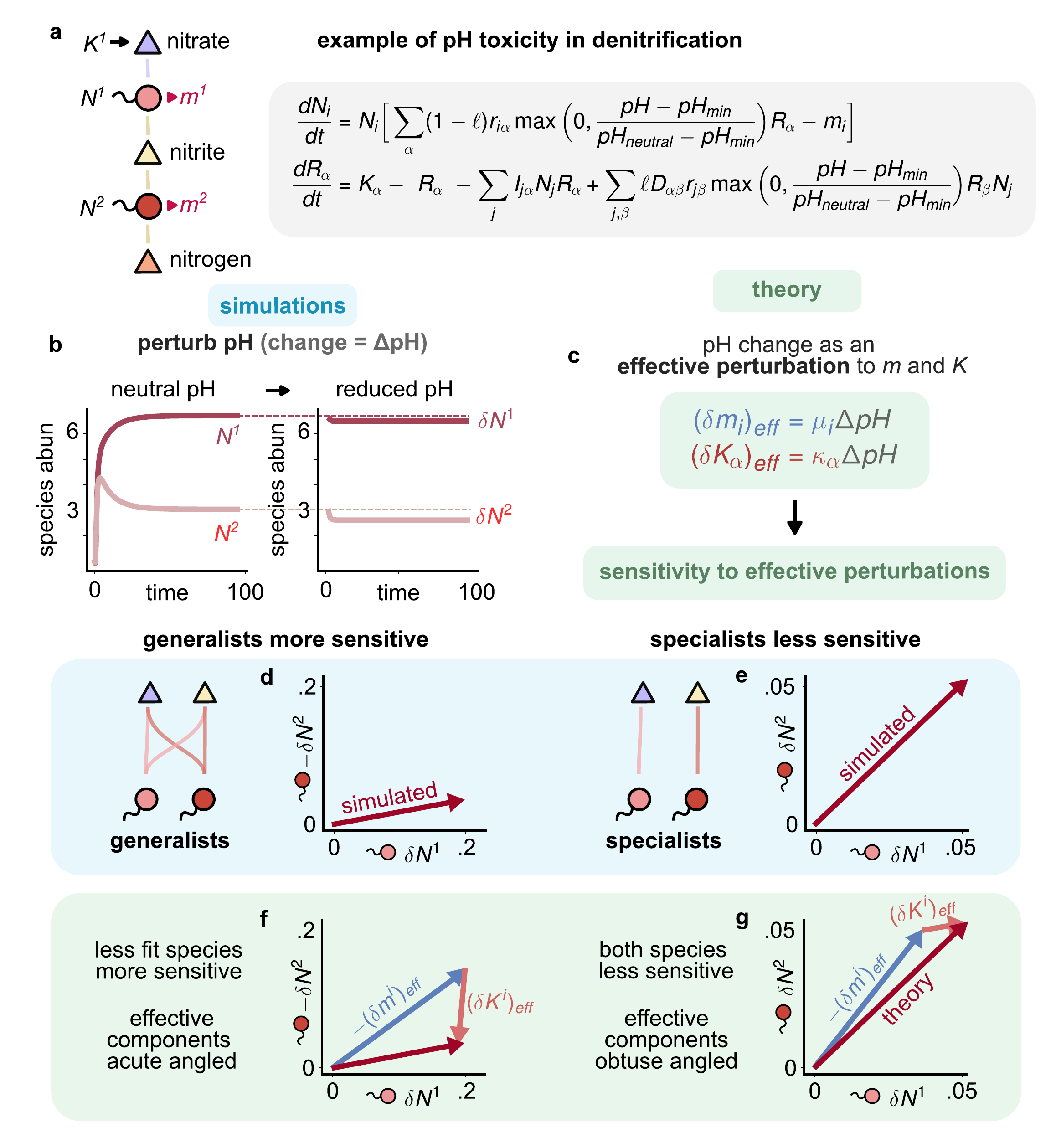}
    \caption{\footnotesize\textsf{\textbf{Emergent geometry of pH sensitivity in bacterial denitrification.
    } 
     (a) Cartoon and equations for bacterial denitrification, containing a chain of conversions from nitrate to nitrite to nitrogen (waste). 
     Reducing pH increases toxicity and reduces growth $r$ of species. The impact/depletion of resources $I$ is not affected by pH~\cite{crocker2023genomic}. No species can use nitrogen.
     (b) Changes in species steady states upon perturbing the pH for (d) an ecosystem of two generalists that can both consume nitrate and nitrite and (e) an ecosystem composed of two species specializing on either nitrate or nitrite.  
     (c) pH change $\Delta$pH can be mapped to effective mortality $(\delta m)_{\text{\text{eff}}}$ and supply $(\delta K)_{\text{\text{eff}}}$ perturbations, to which our geometric formalism applies.
          \comment{(f--g) Geometry of denitrification: our theoretical predictions (decomposition into effective supply (pink) and mortality (blue) components) agree with results from numerical simulations (d--e) for change in species abundances and explain why generalists are more sensitive to pH perturbations than specialists.}
    }}
    \label{fig:4}
\end{figure*}

\subsection{pH sensitivity in bacterial denitrification}
Our geometric formalism can also be used to analyze global perturbations such as changes to the pH, salinity, or even the introduction of new invasive species into ecosystems. To illustrate this, we focus on understanding the effect of pH changes in denitrifying bacterial communities~\cite{gowda2022genomic, crocker2023genomic}.  Denitrification is the process by which bacteria reduce nitrate to nitrite and ultimately to nitrogen gas in order to extract energy (Fig.~\ref{fig:4}a). Nitrate and nitrite can be used as resources for growth, while the end product nitrogen usually diffuses away as a gas. Recent experiments indicate that there are two distinct types of denitrifying communities: communities of specialists where each step of denitrification is performed by distinct species (i.e., each species can utilize nitrate or nitrite but not both) and communities consisting of generalists where each species can use both nitrate and nitrite as an energy source~\cite{crocker2023genomic}. These same experiments show that reducing pH decreases growth from consuming nitrite without affecting how species impact/deplete it. The dynamics of these communities can be summarized by equations of the form
\begin{widetext}
\begin{equation}
\begin{aligned}
\dv{N_i}{t} &= N_i\qty[\sum_\alpha (1-\ell) r_{i\alpha}\max\qty(0,\frac{\text{pH} - \text{pH}_{\text{min}}}{\text{pH}_{\text{neutral}} - \text{pH}_{\text{min}}}) R_\alpha -m_i] \\
\dv{R_\alpha}{t} &= K_\alpha - R_\alpha - \sum_j I_{j\alpha} N_j R_\alpha +  \sum_{j,\beta} \ell D_{\alpha\beta}r_{j\beta}\max\qty(0,\frac{\text{pH} - \text{pH}_{\text{min}} }{ \text{pH}_{\text{neutral}} - \text{pH}_{\text{min}}})R_\beta N_j.
\end{aligned}\label{Eq:Denit-dyn}
\end{equation}
\end{widetext}
where $\alpha$ runs over the three types of resources: nitrate, nitrite, and nitrogen, $r_{i\a}$ are species' growth preferences at neutral pH, $I_{j\a}$ are the impact coefficients unchanged by pH, and $D_{\a\b}$ is a cross-feeding matrix introduced in Sec.~\ref{sec:Mic-CrossFeeding} encoding the denitrification chain. \comment{The $\max$ function captures the effect of changing pH, and enters only in the growth and production terms containing $r_{i\a}$, not the depletion (impact) terms containing $I_{i\a}$.} For specialist bacterial species $i$, $r_{i\a}$ is non-zero for either nitrites or nitrates whereas generalist species have non-zero growth coefficients for both these metabolites (Fig.~\ref{fig:4}). 

We wanted to understand the differences between how generalist and specialist communities respond to pH changes of the form  ${\text{pH}_{\text{neutral}} \rightarrow \text{pH}_{\text{neutral}} + \Delta \text{pH}}$. Following Sec.~\ref{sec:GCRM-intro}, near the steady state, $\vec{R}^*$ and $\vec{N}^*$, we can define generalized interactions for species growth $C_{i\a}$, impact $E_{i\a}$, resource interactions $Q_{\a\b}$, as well as the generalized projector $P^\a_\b$ as in Eq.~\eqref{Eq:generalCEQ}. As shown in Appendix~\ref{Appendix:GCRM-LinResp}, to linear order the pH perturbation $\Delta \text{pH}$ can be written as effective perturbations to resource supply $\d K_\a^{\text{eff}}$ and species mortality $\d m_i^{\text{eff}}$ of the form (Fig.~\ref{fig:4}c):
\begin{equation}
\begin{aligned}
\d m_{i}^{\text{eff}} &= \sum_\a {(1-\ell)r_{i\a} R^*_\a \over \text{pH}_{\text{neutral}} - \text{pH}_{\text{min}}} \Delta \text{pH} \\
\d K_\a^{\text{eff}} &= \sum_{j,\beta} {\ell D_{\a\b} r_{j\b} R^*_\b N^*_j \over \text{pH}_{\text{neutral}} - \text{pH}_{\text{min}}} \Delta \text{pH}.
\end{aligned}\label{Eq:Denit-EffPert}
\end{equation}
Notice that the effective perturbations depend on the steady-state abundances . The change in species abundances, $\d N^i$, and resource abundances, $\d R^\alpha$, must satisfy Eq.~\eqref{Eq:GCRM-lin} except with the effective perturbations defined above,
\begin{align}
C_{i \a} \d R^\a &= \d m_{i}^{\text{eff}} \nonumber \\
Q_{\a \b} \d R^\b  + E_{i \a} \d N^i &= \d K_\a^{\text{eff}}.
\label{Eq:Denit-lin}
\end{align}

To understand the effect of lowering pH, we simulated the ecological dynamics of both specialist and generalist communities using Eq.~\eqref{Eq:Denit-dyn}. We first let the communities reach a steady-state at neutral pH and then reduced the pH slightly until a new a new steady state was reached (Fig.~\ref{fig:4}b) (see Appendix ~\ref{Appendix:Methods} for details). We used these simulations to calculate the change in the steady-state species and resource abundances $\d N^i$ and $\d R^\a$ at neutral and low pH. We then compared the results with theoretical predictions made using Eqs.~\eqref{Eq:Denit-lin} and~\eqref{Eq:Denit-EffPert} (Fig.~\ref{fig:4}f). As can be seen in Figs.~\ref{fig:4}d-g, our theory agrees with results from numerical simulations.

\begin{figure*}[ht!]
    \centering
    \includegraphics[width=0.6\textwidth]{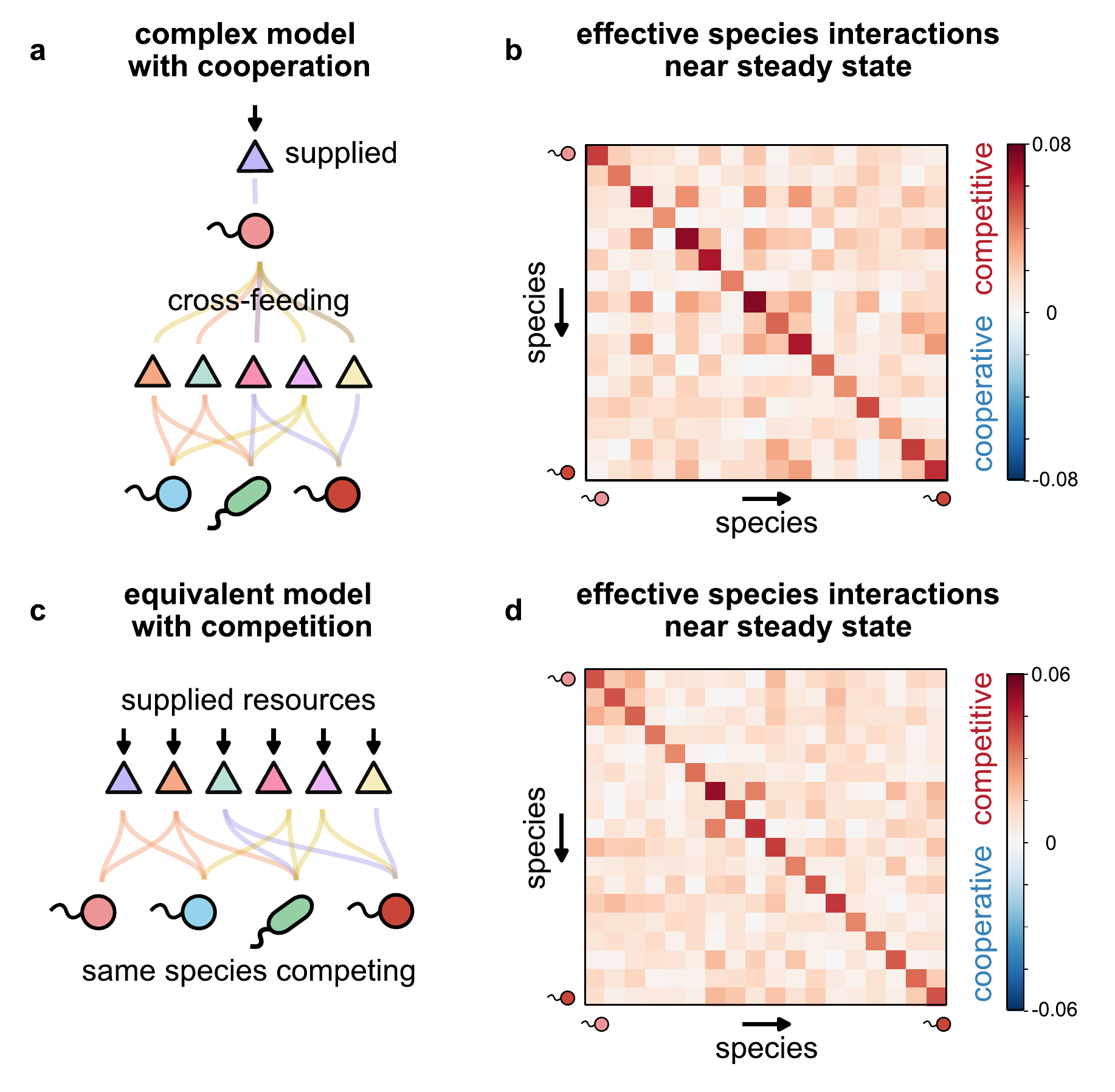}
    \caption{\footnotesize\textsf{\textbf{Disentangling cooperation from competition in complex ecological models
    } 
     \comment{(a) Cartoon of dynamics of a non-linear cross-feeding model seeded with 100 species and 51 resources, where only one resource is externally supplied and all others are cross-fed. 
     (c) Cartoon of equivalent competitive model where all resources are now supplied externally and there is no cross-feeding (see Appendix \ref{Appendix:Methods})
(b)--(d) We simulate both models and compute the effective species interaction matrix $A_{ij} = [CQ^{-1}E^T]_{ij}$ (see Eqs. (\ref{Eq:AijCrossFeeding}) and \ref{Eq:AijLinResourceModel}) for surviving species in both models near steady-state. Positive values (red) indicate competitive while negative (blue) indicate cooperative interactions. The interaction matrix for both models is consistent with competition. The species interaction matrix for cross-feeding masks cooperation, even though the entire community survives due to it.}
    }}
    \label{fig:5}
\end{figure*}

As can be seen in Figs.~\ref{fig:4}d-g, generalist and specialist communities respond qualitatively differently to changes in pH. In the community of generalists, one of the two species experiences a much larger change in steady-state abundance than the other. In contrast, in the specialists community both species experience a similar change in abundance. \comment{These results also hold for randomly generated generalists and specialists (Fig. \ref{fig:S2}).} Our geometric framework also allows us to understand the origin of this difference by decomposing $\d N^i$  into changes in the naive abundance,  $\d K^i$, and changes in the mortality burden, $\d m^i$ (Figs.~\ref{fig:4}f-g, pink and blue respectively). In communities of generalists, both supply and mortality components are acute angled, while in the community of specialists, they are obtuse angled. As discussed previously, the angles between components are a consequence of non-reciprocity in species-resource interactions. The more non-reciprocal the effective interactions, the greater the angle between components. Specialist communities divide labor and have greater effective non-reciprocity. Further, the two components of the pH sensitivity vector are somewhat negatively correlated for generalists and positively correlated for specialists. The vector sum of these components is skewed towards one species for generalist communities, but of similar magnitude for specialist communities. The two different orientations of $\d K^i$ (downward for generalists and upward for specialists) simply reflect that the effective resource supply is increased for both species in specialist communities whereas it is decreased due to competition in generalist communities. This difference is the primary driver of the greater sensitivity of generalist communities to changes in pH. These results are consistent with recent experimental observations on specialist and generalist communities~\cite{crocker2023genomic}. 

\comment{\section{Advantages of universal niche geometry}
\label{sec:advNicheGeom}
In addition to revealing the geometric structure of solutions to generalized CRMs (Eq. (\ref{Eq:generalRD})), our framework also reveals several biological insights. In this section, we highlight a few such selected insights which go beyond ``solving the models'' or interpreting the terms.

\textbf{Correspondence between simple and complex dynamics:} First, our framework shows a correspondence between arbitrarily complex ecological models and equivalent simple MacArthur models. Any complex ecological model with dynamics of the form of Eq. (\ref{Eq:generalRD}) is generally not solvable. But around steady state, we show any such model can be approximated by a simple MCRM whose linearized dynamics are equivalent to the complex models. This resulting MCRM is exactly solvable (its solutions being Eq. (\ref{Eq:GCRM-SS-final}). This correspondence between complex ecological dynamics and MCRMs highlights why MCRMs, despite their simplifying assumptions, are the \textit{de facto} general models of ecological dynamics.

\textbf{Disentangling competition from cooperation:} Second, a powerful consequence of this correspondence is that it is almost impossible to learn the true biological processes underlying the ecological dynamics of a community perturbed around steady state. The latter is common practice in ecological experiments and field studies. The reason it is hard to learn the true mechanism is that there are infinitely many models with different processes and dynamics that have the same linearized behavior and thus map to the same effective MCRM. A strong result of this is that it is impossible to disentangle competition from cooperation using perturbation experiments of community dynamics (Fig. \ref{fig:5}a--d). Near steady state, a community that coexists only due to cross-feeding will behave indistinguishably from a community that coexists purely through nutrient competition (Fig. \ref{fig:5}b and \ref{fig:5}d). In the language of effective interactions, both communities will have effective species interaction matrices $A$ that will be consistent with competition ($A_{ij} > 0$) and will mask any mutualisms. Even though both interaction matrices have different decompositions (Appendix \ref{Appendix:GCRM-LinResp}), what is observed in the dynamics are not the decompositions, but the full composite quantities.

\textbf{Defining decoupled niches:} Finally, another powerful consequence of the steady-state equivalence between several ecological models is that corresponding to every interacting ecological system, there is an equivalent non-interacting system. This is easiest to see the language of linearized niche geometry. The benefit of the linear algebraic solutions is that one can go the eigenbasis of the linear transformations we have identified, which diagonalizes the interactions and therefore decouples species from each other. In this way, one can rigorously see and define niches. Niches are eigenvectors of $ACQ^{-1}$, which represent combinations of resources, each of which dominantly contributes to the abundance  of exactly one species. Each combination of resources is thus each species' niche. People have long had the idea that when species are generalists, it is complex combinations of resources that comprise niches, but thus far there have been no explicit way of computing them. The inherent linear algebraic nature of our framework provides a clear way to compute niches as eigenvectors of the effective transformations we have identified. Indeed, this can be done for any complex ecological dynamics (including community data) as long as we have enough samples to learn the matrix $ACQ^{-1}$. Crucially, knowing niches does not require knowing all the complex ecological dynamics of a community.
}

\section{Discussion}

In this paper, we developed a theory for the steady-state response of ecosystems subject to environmental perturbations. Our theory characterizes the effect of perturbations in resource supply and species mortality on changes in the steady-state species and resource abundances. We show that our framework applies to a variety of generalized consumer-resource models, including models with nonlinear growth and impact (see Appendix ~\ref{Appendix:GCRM-LinResp}). We also show how our theory can be applied to a wider class of environmental perturbations such as pH changes by recasting such changes as effective perturbations to resource supply and mortality. \comment{We conclude by showing the explicit conceptual advantages that our theory offers over simply simulating the underlying ecological models.} For this reason, our framework represents a powerful framework for characterizing how ecosystems respond to environmental perturbations.

The rich mathematical structure of our theory reveals a natural geometry that underlies consumer-resource models.  Ecological variables -- species and resource abundances and fluxes -- live in four distinct vector spaces that are related by generalized susceptibilities. This allows us to decompose the effect of environmental perturbations in terms of  canonical vectors whose geometric relationship encodes information about competition and niche structure. These vectors represent natural ``emergent'' variables that characterize steady states and linear response. These observations suggest that our geometric framework may be useful for defining coarse-grained collective ecological modes~\cite{shan2023annotation, moran2022defining}.

Our theory focuses on describing how an ecosystem's steady state changes in response to environmental perturbations. In the future, it will be interesting to see if our framework can also provide geometric insight into dynamical phenomena, e.g., the onset of instability and chaos in diverse ecosystems with non-reciprocal interactions~\cite{blumenthal2023phase}. There is a long history of explaining such dynamical phenomena using geometry~\cite{strogatz2018nonlinear}. Recent work using DMFT to analyze chaos in generalized Lotka-Voltera models suggests that abundant species  reach long-lived transients that share many properties with steady states~\cite{de2023many}. If this is also true in CRMs, then it may be possible to adapt the geometric framework developed here to describe the dynamics of abundant species and resources even in the chaotic phase.

We are also interested in extending our framework to explore the interplay between ecological and evolutionary dynamics. The introduction of a new mutant can be viewed as a small perturbation to the environment of the remaining species~\cite{good2018adaptation, mahadevan2023spatiotemporal}. For this reason, it maybe possible for some types of eco-evolutionary  dynamics (e.g., strong selection, weak mutation) to be analyzed using the methods outlined here. Finally, we note that the basic geometric and mathematical structure outlined here also seems to be present in other complex systems that can be described using two distinct sets of variables analogous to species and resources, including mechanical metamaterials~\cite{rocks2022integrating} and linear regression~\cite{rocks2021geometry}. This suggests that many of the ideas presented here maybe applicable to complex systems beyond ecological settings.

\vskip 10pt

\textbf{Acknowledgements.}
We thank Zhijie (Sarah) Feng and Emmy Blumenthal for discussions. This work was funded by NIH NIGMS R35GM119461 to P.M. and Chan-Zuckerburg Institute Investigator grant to P.M. A.G. acknowledges support from the Gordon and Betty Moore Foundation as a Physics of Living Systems Fellow under grant number GBMF4513, the Ashok and Gita Vaish Junior Researcher Award, as well as the Government of India's DBT Ramalingaswami Fellowship.

\bibliography{bibliography}

\clearpage
\appendix

\renewcommand{\thefigure}{S\arabic{figure}}
\setcounter{figure}{0}
\section{Relating the Generalized MacArthur Consumer-Resource Model to Lotka Volterra} \label{Appendix:GLVderivation}
In this appendix, we will show that when resource dynamics of a Generalized MCRM reach a steady state, the corresponding steady-state species abundances can be obtained using an equivalent Generalized Lotka-Volterra (GLV) model. This statement is always true for Generalized MCRMs at steady state, regardless of the pace of resource dynamics relative to species dynamics. Setting resources to steady state, ${\dv{R_\a}{t} = 0}$, we get the following equation in terms of steady-state resource abundances $\vec{R}^*$:
\begin{equation}
\begin{aligned}
0 &= R^*_\alpha \qty(K_\alpha - \sum_\beta Q_{\alpha \beta} R^*_\beta) - \sum_{j} E_{i \alpha} N_j R^*_{\alpha},\\
0 &= K_\a - \sum_\beta Q_{\a\b} R^*_\b - \sum_j E_{j\a} N_j.
\end{aligned}
\end{equation}

Solving for $R*_\b$ and substituting into the species dynamics Eq.~\eqref{Eq:GMCRM}, we get
\begin{align}
\dv{N_i}{t} &= N_i \qty[ \sum_\a C_{i\a} Q^{-1}_{\a\b} \qty(K_\a - \sum_j E_{j\a} N_j ) - m_i ],
\end{align}
which can be rearranged to get the corresponding GLV equations for consumer species
\begin{align}
\dv{N_i}{t} &= N_i \qty( \mathcal{K}_i - \sum_j A_{ij} N_j ),
\end{align}
 where we have defined the ``effective''  species-species interaction matrix 
\begin{align}
A_{ij}&= \sum_{\alpha,\beta} C_{i \alpha} Q^{-1}_{\alpha \beta} E_{j \beta}
\end{align}
and written the effective carrying capacity of a species as the difference
\begin{align}
\mathcal{K}_i &=\sum_{\alpha, \beta} C_{i \alpha} Q^{-1}_{\alpha \beta}  K_\beta - m_i
\end{align}
between the ``naive'' fitness (first term) and the mortality rate $m_i$.

\section{Steady state Abundance and Resource Expressions} \label{Appendix:SSExpression}

In this appendix, we will solve for the steady-state species and resource abundances in Generalized MacArthur Consumer-Resource Models (GMCRMs). For any number of initially added species $S$ and resources $M$, we will solve for the abundances of the $M^*$ non-depleted resources and the $S^*$ surviving species using the linear equations
\begin{equation}
\begin{aligned}
0 &=C^* \vec{R}^* - \vec{m}^*\\
0 &= \vec{K}^*-Q^* \vec{R}^* - E^{*T}\vec{N}^*,
\end{aligned}
\end{equation}
where we have introduced the $S^*$-dimensional vectors of species abundances of surviving species $\vec{N}^*$ and their corresponding mortality rates or maintenance costs $\vec{m}^*$, and the $M^*$-dimensional vectors of resource abundances for non-depleted resources $\vec{R}^*$ and their corresponding carrying capacities $\vec{K}^*$. In what follows, we will drop the $^*$ for simplicity. We can write this as a block matrix equation of the form
\be
\mqty[0 & C^*\\
E^{*T} & Q]
\mqty[\vec{N}^*\\
\vec{R}^* ]
=
\mqty[\vec{m}^*\\
\vec{K}^*].
\label{Eq:MacArthurBlockM}
\ee

In what follows we will assume $Q=I$. A similar derivation can be done for a general $Q$, with solutions as in Eq. (\ref{Eq:GMCRM-SS-final}) of the main text. To solve this equation, we use standard block-diagonal formulas (using the Schur complement of the lower diagonal matrix $I$),
\be
\mqty[
0 & C^*\\
E^{*T} & I
]^{-1}
\\=
\mqty[
-(C^*  E^{*T})^{-1} & (C^*  E^{*T})^{-1} C^*\\
E^{*T}(C^*  E^{*T})^{-1}& I- E^{*T}(C^*  E^{*T})^{-1} C^*
].
\ee
Note that since $S^*\le M^*$, we know that  $(C^* E^{*T})$ is full rank and therefore invertible.
To simplify notation, notice that $E^{*T}(C^* E^{*T})^{-1}C^*$ is a (oblique) projector
onto the $S^*$-dimensional subspace of the full $M^*$-dimensional resource space spanned by the surviving species.
To see this, we define
\be
P_*=E^{*T}(C^*  E^{*T})^{-1}C^*,
\ee
which displays the projector property
\be
P_*^2=P_*.
\ee
We can also define a ``species-interaction'' matrix 
\be
A^*=C^*  E^{*T},
\ee
in terms of which we can write the projector as
\be
P_*=E^{*T}A^{* -1}C^*.
\ee

In terms of these new matrices, we have
\be
\mqty[
0 & C^*\\
E^{*T} & I
]^{-1}
=
\mqty[
-A^{* -1} & A^{* -1} C^*\\
E^{*T}A^{* -1}& I- P_*
],
\ee
which gives the required steady-state species and resource abundances $N^*$ and $R^*$, respectively, as
\be
\mqty[
\vec{N}^*\\
\vec{R}^* 
]
=
\mqty[
-A^{* -1}\vec{m}^* + A^{* -1} C^*\vec{K}^* \\
E^{*T}A^{* -1}\vec{m}^*+(I- P_*)\vec{K}^*
].
\label{Eq:AppBSSExp}
\ee

\comment{\section{Operators in Einstein Notation} \label{Appendix:EinsteinOperators}

The transformations in Fig.~\ref{fig:1}b allow us to ``transport'' vectors between these four vector spaces. For example, the resource abundance vector $R^\alpha$ has three natural counterparts in resource flux space, species flux space, and species abundance space given by 
\begin{equation}
\begin{aligned}
R_\alpha &= Q_{\alpha \beta}R^\beta  \\
R_i &= C_{i \beta}R^\beta  \\
R^i &= A^{ij}R_j = A^{ij}  C_{j \beta}R^\beta .
\end{aligned}\label{Eq:4Rvec}
\end{equation}
One can perform similar constructions for the resource supply rates $K_\alpha$,
\begin{equation}
\begin{aligned}
K^\alpha &= Q^{\alpha \beta}K_\beta  \\
K_i &= C_{i \alpha}K^\alpha=  C_{i \alpha} Q^{\alpha \beta}K_\beta \\
K^i &= A^{ij}K_j  =A^{ij} C_{j \alpha} Q^{\alpha \beta}K_\beta, 
\end{aligned}\label{Eq:4Kvec}
\end{equation} 
the steady-state species abundances $N^i$,
\begin{equation}
\begin{aligned}
N_i &= A_{ij}N^j  \\
N_\alpha &= E_{j \alpha}N^j  \\
N^\alpha &=Q^{\alpha \beta}N_\beta= Q^{\alpha \beta}E_{j \beta}N^j,
\end{aligned}\label{Eq:4Nvec}
\end{equation}
and the species mortality rates $m_i$,
\begin{equation}
\begin{aligned}
m^i &=A^{ij} m_j  \\
m_\alpha &= E_{j \alpha} m^j = E_{j \alpha}A^{jk} m_k \\
m^\alpha &= Q^{\alpha \beta}m_\beta= Q^{\alpha \beta}E_{j \beta} m^j=E_j^\alpha m^j.
\end{aligned}\label{Eq:4mvec}
\end{equation}

Collectively, Eqs.~\eqref{Eq:4Rvec}-\eqref{Eq:4mvec} define sixteen vectors providing a complete characterization of the ecological steady-state properties of Generalized MacArthur CRMs. In Fig.~\ref{fig:1}c, we graphically illustrate each of these vectors, along with a brief ecological interpretation. 
}

\section{Susceptibilities of Generalized MacArthur CRMs} 
\label{Appendix:MCRM-susc}

Using Appendix~\ref{Appendix:SSExpression}, we can also calculate the matrix form of the susceptibilities, as we show in this appendix. For $Q=I$, we see by differentiating Eq.~\eqref{Eq:MacArthurBlockM}, we get
\be
\mqty[
0 & C^*\\
E^{*T} & I
]
\mqty[
\pdv{\vec{N}^*}{\vec{m}^*} & \pdv{\vec{N}^*}{\vec{K}^*}\\
\pdv{\vec{R}^*}{\vec{m}^*} & \pdv{\vec{R}^*}{\vec{K}^*}
]
=
\mqty[
I & 0\\
0 & I 
]
\ee
so that we have
\be
\mqty[
\pdv{\vec{N}^*}{\vec{m}^*} & \pdv{\vec{N}^*}{\vec{K}^*}\\
\pdv{\vec{R}^*}{\vec{m}^*} & \pdv{\vec{R}^*}{\vec{K}^*}
]
=
\mqty[
-A^{* -1} & A^{* -1} C^*\\
E^{*T}A^{* -1}& I- P_*
].
\ee
This also allows us to write the steady-state abundances in terms of the susceptibilities as  
\be
\mqty[
N^*\\
R^* 
]
=
\mqty[
\pdv{\vec{N}^*}{\vec{m}^*} & \pdv{\vec{N}^*}{\vec{K}^*}\\
\pdv{\vec{R}^*}{\vec{m}^*} & \pdv{\vec{R}^*}{\vec{K}^*}
]
\mqty[
m^*\\
K^* 
]\\
=
\mqty[
\pdv{\vec{N}^*}{\vec{m}^*} m^*+\pdv{\vec{N}^*}{\vec{K}^*}K^*\\
\pdv{\vec{R}^*}{\vec{m}^*} m^*+\pdv{\vec{R}^*}{\vec{K}^*}K^*
].
\ee

Thus, we see that each of the terms in the steady-state expression in Eq.~\eqref{Eq:AppBSSExp} involves these susceptibilities, and these susceptibilites contain all the relavant ecological interactions.

\comment{\section{Calculating steady states of Generalized MacArthur CRMs using Einstein notation} 
\label{Appendix:MCRM-EinNot}
As shown in the main text, we can also express all ecological quantities in Einstein notation. In this appendix, we will demonstrate how to solve the GMCRMs steady-state conditions, written in Einstein notation as
\begin{equation}
\begin{aligned}
0 &= C_{i\beta} R^\beta - m_i\\
0 &= K_\alpha-Q_{\alpha\beta} R^\beta - E_{j\alpha}N^j,
\end{aligned}
\end{equation}
to obtain the steady-state expression in Eq.~\eqref{Eq:GMCRM-ess}, again written in Einstein notation as
\begin{equation}
\begin{aligned}
N^i &=K^i -m^i\\
R^\alpha& =m^\alpha + (\delta^\alpha_\beta -P^\alpha_\beta) K^\beta.
\end{aligned}
\end{equation}

First, we derive the steady-state species abundances $N^i$ from the steady-state conditions,
\begin{equation}
\begin{aligned}
 E_{j\alpha}N^j &= K_\alpha - Q_{\alpha\beta} R^\beta\\
 Q^{\gamma \alpha}E_{j\alpha}N^j &= Q^{\gamma \alpha}K_\alpha - Q^{\gamma \alpha}Q_{\alpha\beta} R^\beta\\
E_j^\gamma N^j &= K^\gamma - R^\gamma \\
C_{k \gamma}E_j^\gamma N^j &= C_{k \gamma}K^\gamma - C_{k \gamma}R^\gamma \\
C_{k \gamma}E_j^\gamma N^j &= C_{k \gamma}K^\gamma - m_k\\
 A_{kj} N^j &= K_k - m_k\\
 A^{ik}A_{kj} N^j &= A^{ik}K_k - A^{ik}m_k\\
N^i &= K^i - m^i.
\end{aligned}
\end{equation}

Using this result, we can then derive the steady-state resources abundances $R^\alpha$ from the steady-state conditions. Starting from the third line in the previous derivation,
\begin{equation}
\begin{aligned}
R^\alpha  &= K^\alpha - E_j^\alpha N^j\\
&= K^\alpha - E_j^\alpha (K^j - m^j)\\
&= m^\alpha  + K^\alpha - E_j^\alpha A^{jk}C_{j\beta}K^\beta\\
&= m^\alpha  + K^\alpha - E_j^\alpha C^j_\beta K^\beta\\
&= m^\alpha + \qty(\delta^\alpha_\beta - P^\alpha_\beta)K^\beta
\end{aligned}
\end{equation}}

\section{Calculating angles between geometric components}
\label{Appendix:Geom-Angles}
In this appendix, we will calculate the angle between the components of steady-state resource and species abundances in Generalized MacArthur Consumer-Resource Models (GMCRMs). Specifically, we will show that the magnitude of the angle depends on the degree of non-reciprocity in the ecological interactions. Due to the mathematical equivalence between steady-state abundances in GMCRMs and sensitivity to perturbations in Generalized Consumer-Resource Models, the angles derived here also generalize to those models, albeit they reflect the non-reciprocity in the generalized interactions described in the main text.

We will start by noting the steady-state expressions for species and resources in GMCRMs,
\begin{equation}
\begin{aligned}
\vec{N}& = A^{-1} [CQ^{-1} \vec{K} -\vec{m}] \\
\vec{R}& = Q^{-1} [E^T A^{-1} \vec{m} + (1-P) \vec{K}].
\label{Eq:AppGMCRM-SS}
\end{aligned}
\end{equation}
It will suffice to show the angle between the two components of the steady-state abundances first. Consider the case when $\vec{m}$ and $\vec{K}$ are of equal magnitude, as is the case throughout our manuscript. Relaxing this assumption will generally change the expression for the angles, requiring us to incorporate relative changes in mortality and supply. 

When mortality and supply are comparable, the cosine of the angle $\theta_N$ between the two components of the steady-state species abundance vector will be proportional to the inner product of the equivalent vector components as 

\begin{align}
\cos \theta_N \propto \langle A^{-1} | A^{-1} C Q^{-1} \rangle,
\label{Eq:Angle-Metrics-N}
\end{align}
where the full expression for $\cos\theta_N$ will also involve the norms of the matrices involved. By straightforward inspection, we can see that $\theta_N$ is close to zero when the inner product is close to 1, which corresponds to the matrices $A^{-1}$ and $Q^{-1}$ being approximately diagonal. Note that these matrices are the generalized metrics from our geometric theory. With increasing non-reciprocity, both metrics $A^{-1}$ and $Q^{-1}$ will increasingly gain asymmetric off-diagonal entries. This will decrease the inner product, and thus increase the angle $\theta_N$ as per Eq.~\eqref{Eq:Angle-Metrics-N}. 

Similarly, the angle $\theta_R$ between the two components of the steady-state resource abundance vector will be 
\begin{align}
\cos \theta_R \propto \langle Q^{-1} E^T A^{-1} | (I - E^T A^{-1} C Q^{-1} \rangle.
\label{Eq:Angle-Metrics-R}
\end{align}
Similar to $\theta_N$, this angle also increases as the metrics $A^{-1}$ and $Q^{-1}$ become increasingly asymmetric. Due to the mathematical equivalence between the steady-state expressions in GMCRMs and linear response expressions in generalized nonlinear consumer-resource models, the geometric intuition presented here is preserved. Thus, the angle between geometric components generically indicates the extent of non-reciprocity in an ecosystem.

\section{Calculating linear response for common variants of consumer-resource models}
\label{Appendix:GCRM-LinResp}
In this appendix, we will calculate the generalized ecological interactions for common variants of consumer-resource models. We will focus on models where all interactions are resource mediated [as in Eq.~\eqref{Eq:generalRD}], since this is where our niche geometry is valid.

\subsection{Linear Resource Model}
For reference:
\begin{equation}
\begin{aligned}
C_{i\a} &= \pdv{g_i}{R^\alpha}\eval_{\vec{R}^{*}}\\
E_{i\a} &= \pdv{f_\a}{N^i}\eval_{\vec{R}^*, \vec{N}^{*}} \\ 
Q_{\a\b} &= \pdv{f_\a}{R^\b}\eval_{\vec{R}^*, \vec{N}^{*}}+ \pdv{q_\a}{R^\b}\eval_{\vec{R}^*},
\end{aligned}
\end{equation}
\begin{equation}
\begin{aligned}
\dv{N_i}{t} &= N_i\qty(g_i(\vec{R}) - m_i) \\
\dv{R_\alpha}{t} &= h_\alpha(\vec{K}) - q_\alpha(\vec{R}) - f_{\alpha}(\vec{R}, \vec{N}),
\end{aligned}
\end{equation}

We start with the linear resource model with 
\begin{equation}
\begin{aligned}
\dv{N_i}{t} &= N_i g_i\qty(\sum_\alpha \mathrm{w}_{i \alpha} c_{i \alpha} R_\alpha-m_i)\\
\dv{R_\alpha}{t} &= {K_\alpha } - \omega R_\alpha - \sum_j N_j e_{j \alpha}R_\alpha
\end{aligned}
\end{equation}

For future, reference it will be helpful to define the community-level functional impact
\be
f_\alpha = \sum_i e_{i \alpha} N^*_i,
\ee
or in Einstein notation
\be
f_\alpha=e_{i \alpha}N^i.
\ee
We can also define the effective degradation rate for resource $\alpha$ as
\be
\omega_\alpha ^{eff}= \omega + f_\alpha.
\ee

Notice this rate only depends on the average population rate through the functional content $f_\alpha$. For this model we have the following generalized interaction matrices:
\begin{equation}
\begin{aligned}
C_{i \alpha} &= g_i \mathrm{w}_{i \alpha} c_{i \alpha}\\
E_{i \alpha} &= e_{i \alpha}R^*_\alpha \\
Q_{\a\b} &= \qty(\omega + f_\alpha)\d_{\a\b} = \omega_\alpha^{eff}\d_{\a\b}.
\end{aligned}
\end{equation}

As a result, we have
\begin{align}
A_{ij} &=[CQ^{-1}E]_{ij} = \sum_{\alpha} {g_i \mathrm{w}_{i \alpha} c_{i \alpha} e_{j \alpha} R^*_\alpha \over \omega_\alpha^{eff}} \label{Eq:AijLinResourceModel}\\
P_\alpha^\beta &= [E^TA^{-1}CQ^{-1}]_{\alpha\beta}= \frac{1}{\omega_\beta^{eff}} \sum_{jk} R^*_\alpha e_{j \alpha}[A^{-1}]_{jk}g_k \mathrm{w}_{k \beta} c_{k \beta} 
\end{align}
Notice that the metric in resource space $Q_{\alpha \beta}$ and the pseudo-metric for the species space $A_{ij}$ only depend on the community population through the effective degradation rates $\omega_\alpha^{eff}=\omega +f_\alpha$ and hence its functional content $f_\alpha$. For the same reason, this is also true of the projector $P$ encoding the metabolic niches occupied by the species in the ecosystem.

\subsection{Microbial Consumer-Resource Model}
\begin{equation}
\begin{aligned}
\dv{N_i}{t} &= N_i g_i\qty(\sum_\alpha (1-l_\alpha) \mathrm{w}_{\alpha} c_{i \alpha} R_\alpha-m_i)\\
\dv{R_\alpha}{t} &= K_\alpha - \omega R_\alpha - \sum_j N_j c_{j \alpha}R_\alpha + \sum_{\beta j} l_\beta {\mathrm{w}_\alpha \over \mathrm{w}_\beta }D_{\alpha \beta}c_{j \beta} R_\beta N_j
\end{aligned}
\end{equation}
Once again the community functional content of resource $\alpha$ as
\be
f_\alpha = \sum_j c_{j \alpha} N^*_j
\ee
and 
\be
\omega_\alpha^{eff} = \omega + f_\alpha.
\ee
It is also helpful to define a production flux due to cross feeding of resource $\alpha$ by species $i$ which is just
\be
\kappa_{i \alpha}^{cf}= \sum_{\beta}  l_\beta {\mathrm{w}_\alpha \over \mathrm{w}_\beta } D_{\alpha \beta} R^*_\beta c_{i \beta}
\ee
and the effective depletion rate of resource $\alpha$ by species $i$:
\be
\kappa_{i \alpha}^{eff} = -c_{i \alpha}R^*_\alpha +\kappa_{i \alpha}^{cf}.
\ee

Straightforward calculation and inspection gives
\begin{equation}
\begin{aligned}
C_{i \alpha} &= g_i \mathrm{w}_{\alpha}(1-l_\alpha) c_{i \alpha} \\
E_{i \alpha} &= -\kappa_{i \alpha}^{eff}  \\
Q_{\alpha \beta} &= (\omega+f_\alpha)\delta_{\alpha\beta}  -  l_\beta {\mathrm{w}_\alpha \over \mathrm{w}_\beta }D_{\alpha \beta}f_\beta.
\end{aligned}
\end{equation}

Notice, by definition $C_{i \alpha}$ is just the marginal growth rate of species $i$ if one varies resource $\alpha$, so we will denote it
\be
\Delta g_{i \alpha} \equiv C_{i \alpha} = g_i \mathrm{w}_{\alpha}(1-l_\alpha) c_{i \alpha}.
\ee
Furthermore, we can define the total production  flux of resource $\alpha$ due to crossfeeding as
\be
\kappa_\alpha^{cf}= \sum_j \kappa_{j \alpha}^{cf} N^*_j= \sum_{\beta} l_\beta {\mathrm{w}_\alpha \over \mathrm{w}_\beta } D_{\alpha \beta} R^*_\beta f_\beta.
\ee
Notice that we can define how this cross-feeding flux changes as we change $R^*_\beta$ which we can write as
\be
\Delta \kappa_{\alpha \beta}^{cf} = l_\beta {\mathrm{w}_\alpha \over \mathrm{w}_\beta } D_{\alpha \beta} f_\beta.
\ee

With these definitions we have
\begin{equation}
\begin{aligned}
Q_{\alpha \beta} &= \omega_\alpha^{eff} \delta_{\alpha \beta}-\Delta \kappa_{\alpha \beta}^{cf}  \\
A_{ij} &=[CQ^{-1}E]_{ij} = -  \sum_{\alpha\beta} \Delta g_{i \alpha} [Q^{-1}]_{\alpha \beta}
 \kappa_{j \beta}^{eff} \\
P_\alpha^\beta &=[E^TA^{-1}CQ^{-1}]_{\alpha\beta}=  -\sum_{jk\gamma} \kappa_{\alpha j}^{eff} [A^{-1}]_{jk}\Delta g_{k \gamma}[Q^{-1}]_{\gamma \beta}.
\end{aligned}
\label{Eq:AijCrossFeeding}
\end{equation}
Once again notice that the only dependence of the pseudo-metrics $Q$, $A$, and the projector $P$ on the species abundances 
are through $\omega_\alpha^{eff}$ and the total gene content $f_\alpha$.

\comment{
\section{Linear response of the microbial consumer-resource model}
\label{sec:Mic-CrossFeeding}
\begin{figure*}[ht!]
    \centering
    \includegraphics[width=0.7\textwidth]{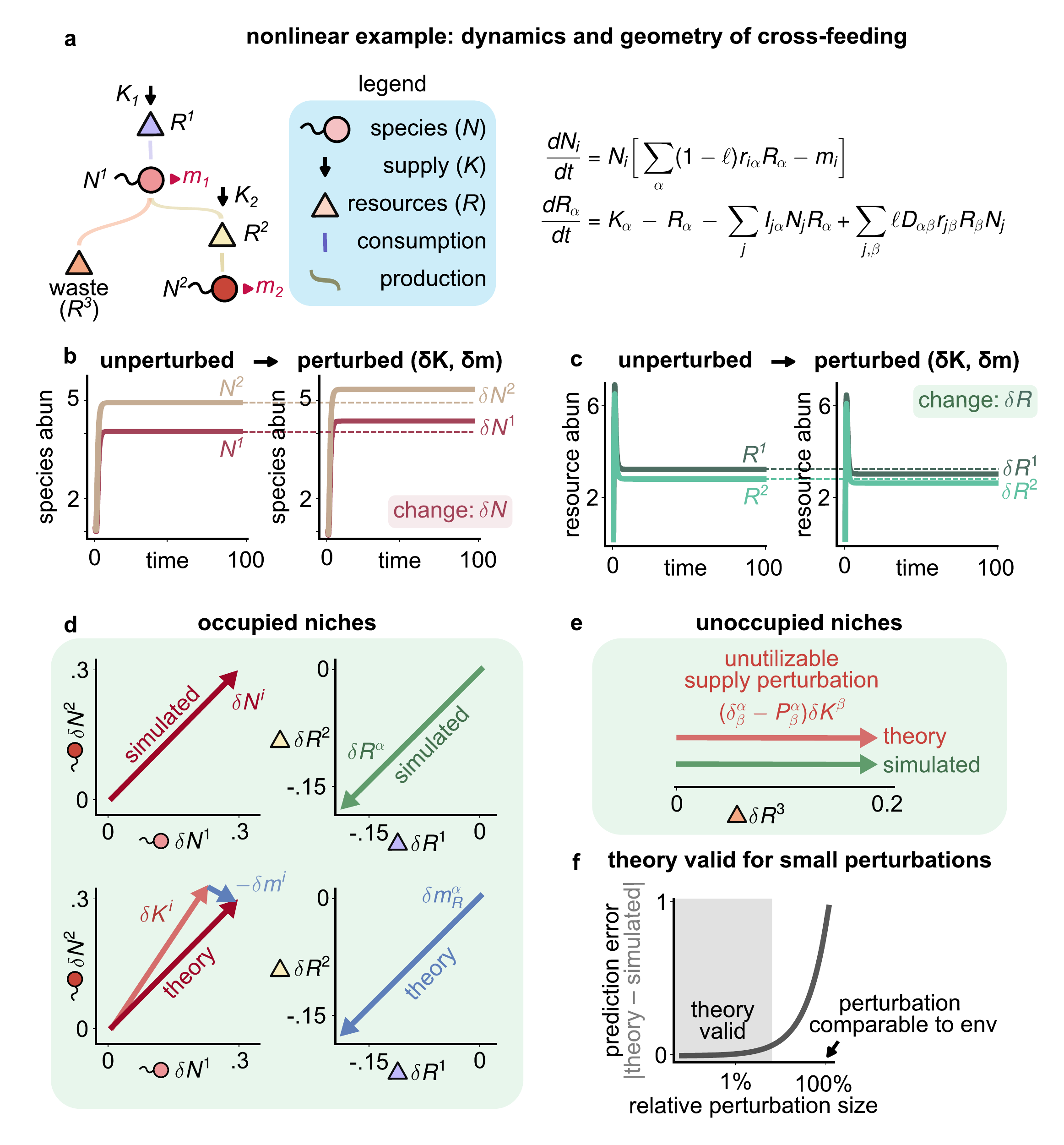}
    \caption{\footnotesize\textsf{\textbf{Decomposing niches in microbial cross-feeding.} 
     (a) Example of dynamics and geometry in a non-linear cross-feeding model with 2 species and 3 resources (with $R^3$ being waste). We perturb the supply $(K_1,K_2)$ to $(K_1',K_2')$ and mortality $(m_1,m_2)$ to $(m_1',m_2')$, and simulate the dynamics to calculate the change in the steady-state  (b) species abundances $\vec{\delta N}$ and (c) resource abundances $\vec{\delta R}$. 
     (d-e) Comparison between results of simulations (red and green vectors) and theoretical predictions (blue and pink vectors).
     (f) Plot of the prediction error of our geometric formalism as a function of the relative perturbation size $\delta K/K$. Predictions agree with simulations for perturbations up to $\approx 10\%$.
     }}
\label{fig:S1}
\end{figure*}

The microbial consumer-resource model has been successfully used to understand a wide variety of experimental observations in microbial ecology~\cite{goldford2018emergent, marsland2019available, marsland2020minimal}. The model describes the secretion of metabolic byproducts by microbes  and was introduced to incorporate the important role played by cross-feeding in microbial ecosystems (Fig.~\ref{fig:S1}a) . The model dynamics take the form of a generalized CRM with equations
\begin{equation}
\begin{aligned}
\dv{N_i}{t} &= N_i \qty[ \sum_{\a=1}^{M^*} (1 - \ell)r_{i\a}R_{\alpha} - m_i] \\
\dv{R_\a}{t} &= K_{\a} - R_{\alpha} - \sum_{j=1}^{S^*}I_{j\a}N_j R_{\a} + \sum_{j,\b} \ell D_{\alpha\beta}r_{j\beta}R_{\beta}N_j,
\label{Eq:MiCRM}
\end{aligned}
\end{equation}
where the cross-feeding matrix $D_{\a\b}$ encodes the conversion of resources $\b$ to metabolic byproducts $\a$, the parameter $\ell$ encodes the fraction of consumed resources that ``leak'' as byproducts, and the consumer and impact preferences are encoded by the matrices $r_{i\a}$ and $I_{j\a}$, respectively. 

For visualization purposes, here we focus on a simple cross-feeding community with 2 species and 3 resources shown in Fig.~\ref{fig:S1}a. We are interested in understanding how this community responds to changes in resource supply and mortality rates (i.e., if we make the change ${K_\a \rightarrow   K_\a + \delta K_\a}$ and the species mortality ${m_i \rightarrow m_i +  \delta m_i}$). We measure the ecosystem's response in two complementary ways: (i) by simulating the dynamics in unperturbed and perturbed environmental conditions  (Figs.~\ref{fig:S1}b-c), and (ii) by calculating the changes $\d N_i$ and $\d R_\alpha$ directly using Eq.~\eqref{Eq:GCRM-ess} (Figs.~\ref{fig:S1}d-e). As shown in Fig.~\ref{fig:S1}f, our analytic predictions can correctly predict the changes in steady states, even for relatively large perturbations of up to $\approx 10\%$.

Our theory also geometrically decomposes the changes in steady states into contributions from supply and mortality perturbations (Figs.~\ref{fig:S1}d-e, pink and blue respectively). The species sensitivity $\d N^i$ is the vector difference between the effective supply perturbation $\d K^i$ (pink) and the effective mortality perturbation $\d m_i$ (blue). Notably, these vectors are roughly orthogonal to each other, suggesting supply and mortality perturbations have qualitatively different effects on both species. In the absence of mortality, changes in both species abundances are positively correlated. But the effect of mortality lowers this correlation by acting in opposite directions on both species. Finally, we note that $\d R^\a$ has only one component, $\d m^\a$, since all resources are utilizable [i.e., $(\d^\a_\b - P^\a_\b)\d K^\b=0$] (see Fig.~\ref{fig:S1}e).

\subsection{Bacterial denitrification}
We start with the bacterial denitrification model summarized as

\begin{widetext}
\begin{equation}
\begin{aligned}
\dv{N_i}{t} &= N_i\qty[\sum_\alpha (1-\ell) r_{i\alpha}\max\qty(0,{\text{pH} - \text{pH}_{\text{min}} \over \text{pH}_{\text{neutral}} - \text{pH}_{\text{min}}}) R_\alpha -m_i] \\
\dv{R_\alpha}{t} &= K_\alpha - \omega R_\alpha - \sum_j I_{j\alpha} N_j R_\alpha  + \sum_{j,\beta} \ell D_{\alpha\beta}r_{j\beta}\max\qty(0,{\text{pH} - \text{pH}_{\text{min}} \over \text{pH}_{\text{neutral}} - \text{pH}_{\text{min}}})R_\beta N_j.
\label{Eq:AppDenit-dyn}
\end{aligned}
\end{equation}
\end{widetext}

For this model, we have the following generalized interactions near a steady state $(\vec{N}^*, \vec{R}^*)$ 
\begin{equation}
\begin{aligned}
C_{i \alpha} &= \sum_\a (1-\ell) r_{i \a} \chi_{\text{pH}} \\
E_{i \alpha} &= { -I_{i\a}R_\a^* + \sum_\b \ell D_{\a\b} r_{i\b} \chi_{\text{pH}} R_\b^* } \\
Q_{\alpha \beta} &= { -\omega - \sum_j I_{j\a}N_j^* + \sum_j \ell D_{\a\b} r_{j\b} \chi_{\text{pH}} N_j^* },
\end{aligned}
\end{equation}
where $\chi_{\text{pH}} = \max\qty(0,{\text{pH} - \text{pH}_{\text{min}} \over \text{pH}_{\text{neutral}} - \text{pH}_{\text{min}}})$. We then have the following generalized linear response equations
\begin{equation}
\begin{aligned}
C_{i \a} \d R^\a &= \d m_{i}^{\text{eff}} \\
Q_{\a \b} \d R^\b  + E_{i \a} \d N^i &= \d K_\a^{\text{eff}},
\end{aligned}
\end{equation}
with the following effective perturbations
\begin{equation}
\begin{aligned}
\d m_{i}^{\text{eff}} &= \sum_\a {(1-\ell)r_{i\a} R^*_\a \over \text{pH}_{\text{neutral}} - \text{pH}_{\text{min}}} \Delta \text{pH} \\
\d K_\a^{\text{eff}} &= \sum_{j,\beta} {\ell D_{\a\b} r_{j\b} R^*_\b N^*_j \over \text{pH}_{\text{neutral}} - \text{pH}_{\text{min}}} \Delta \text{pH}.
\label{Eq:AppEffPertDenit}
\end{aligned}
\end{equation}

This linear response is very similar to cross-feeding in the Microbial Consumer-Resource Model, but with new effective perturbations $\d m_{i}^{\text{eff}}$ and $\d K_{\a}^{\text{eff}}$ described by Eq.~\eqref{Eq:AppEffPertDenit}. Thus the pH change $\Delta \text{pH}$ can be decomposed into effective perturbations in both the mortality and supply. 

Notably, which of the two components dominates depends primarily on the leakage $\ell$, resulting in two distinct regimes. At low leakage $\ell \ll 1$, pH changes act chiefly as mortality changes since $\d m_{i}^{\text{eff}} \gg \d K_{\a}^{\text{eff}}$. At large leakage $\ell \approx 1$, pH changes act chiefly as supply changes $\d m_{i}^{\text{eff}} \ll \d K_{\a}^{\text{eff}}$. As the leakage smoothly increases from 0 to 1, the effective pH perturbations also smoothly shift from mortality-dominated to supply-dominated.}

\section{Simulations and Methods}
\label{Appendix:Methods}

\subsection{MacArthur Consumer-Resource Model}
To simulate the examples in Fig. \ref{fig:2}, we simulated dynamics according to the MacArthur Consumer-Resource Model (MCRM) according to Eqns. (\ref{Eq:GMCRM}), with two  species and two resources ($S=2$ and $M=2$ respectively) and $Q=I$. We simulated two cases: the reciprocal case wth $C=E$ and the non-reciprocal case with $C\neq E$. Both resources were supplied at rates $K_1 = K_2 = 2$, both species experienced mortality at rates $m_1 = m_2 = 0.1$. For $C=E$ (Fig. \ref{fig:2}f), both species were generalists but better at different resources. Species 1 had consumption preferences $C_{11} = 0.3$ and $C_{12}=0.2$, while species 2 had preferences $C_{21}=0.1$ and $C_{22}=0.4$. For $C\neq E$ (Fig. \ref{fig:2}g), species 1 had consumption preferences $C_{11} = 0.5$ and $C_{12}=0.4$ and impacts $E_{11}=0.16$ and $E_{12}=0.01$, while species 2 had preferences $C_{21}=0.4$ and $C_{22}=0.6$ and impacts $E_{21}=0.62$ and $E_{22}=0.72$. In all cases, we numerically simulated the dynamics and obtained the steady-state species and resource abundance vectors (Fig. \ref{fig:2}, blue boxes). To obtain the geometric components from theory (Fig. \ref{fig:2}, green boxes), we computed the two vector components by plugging in the parameters into the solutions in Eqns. (\ref{Eq:GMCRM-SS-final}). We plotted the vector components as well as their vector sum (which represents the predicted steady-state abundance vectors).

\begin{figure*}[ht!]
    \centering
    \includegraphics[width=0.7\textwidth]{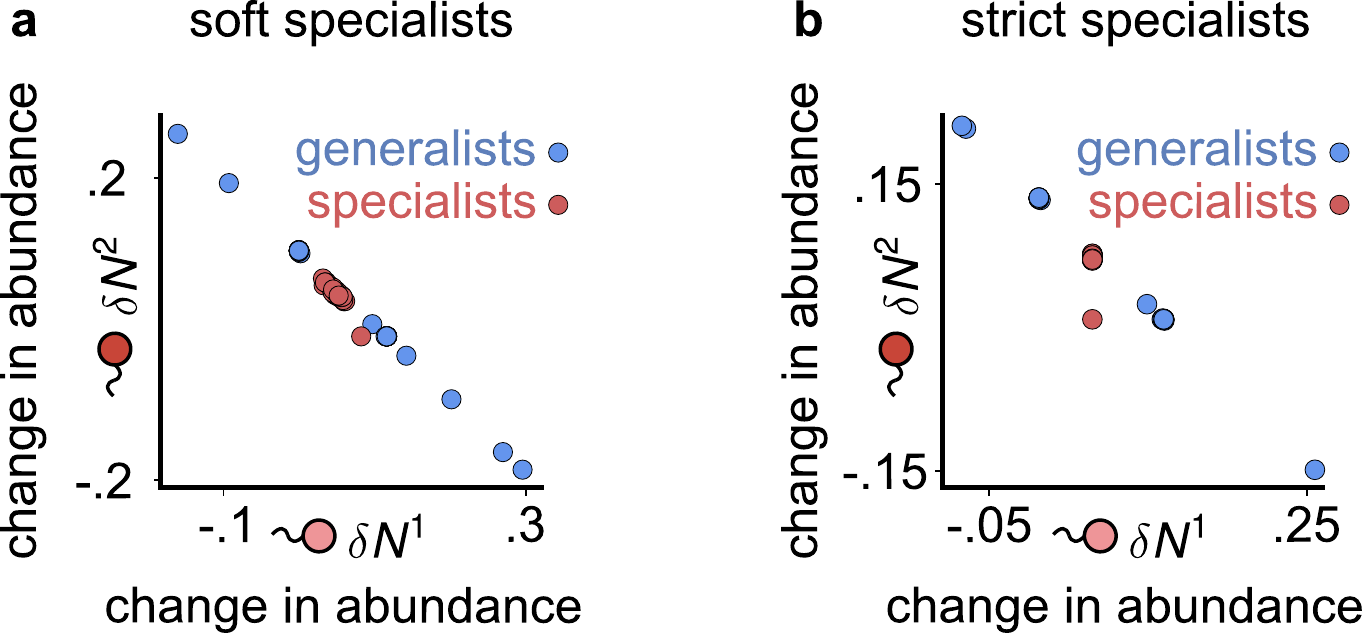}
    \caption{\footnotesize\textsf{\textbf{Generalists differ from specialists in sensitivity to pH perturbations.} \comment{Scatter plot of the sensitivity of two species --- both generalists (blue) or both specialists (red) --- from simulations of the denitrification model with small  pH perturbations. Both specialists tend to cluster around the center of this plot --- indicating that they both have similar sensitivity to pH changes. With two generalists, one of them often shows a much larger sensitivity to pH changes (indicated by concentrating near the extremes). Note that the top-left and bottom-right of the plot contains several nearly overlapping points corresponding to generalists. This observation that the generalist points are often far from the center indicates the skew in their sensitivities --- indicating that one of the generalists is usually much more sensitive than the other.}
     }}
\label{fig:S2}
\end{figure*}

\subsection{Microbial cross-feeding}
To simulate the examples in Fig. \ref{fig:S1}, we simulated dynamics according to the Microbial consumer-resource model with cross-feeding according to Eqns. (\ref{Eq:MiCRM}), with two  species and three resources ($S=2$ and $M=3$ respectively) and leakage $\ell=0.7$. Resources 1 was only supplied externally, while resources 2 and 3 could be produced through cross-feeding, as shown in the cross-feeding network in Fig. \ref{fig:S1}a. We supplied resources 1 and 2 at rates $K_1 = K_2 = 20$, both species experienced mortality at rates $m_1 = m_2 = 1$. No species could consume resource 3 (waste) for growth. Both species were specialists with consumption preferences $r_{11} = 0.9$ and $r_{12}=0$, while species 2 had preferences $r_{21}=0$ and $r_{22}=1$. We assumed that the impacts $I=C$. Starting from random initial conditions with both species and resources present, we numerically simulated the dynamics and obtained the steady-state species and resource abundance vectors (Fig. \ref{fig:S1}b--c). We then perturbed the environment, namely the resource supply and species mortality rates to $K_1=20.05, K_2=19.95, m_1=0.95, m_2=0.95$ and simulated the dynamics to the new steady state with random initial conditions (Fig. \ref{fig:S1}b--c). Using these two steady states (before and after the environmental perturbations), we calculated the species and resource sensitivities, i.e., the change in steady-state species and resource abundances $\d \vec{N}$ and $\d \vec{R}$ (Fig. \ref{fig:S1}d--e). To obtain the geometric components of these sensitivities from theory (Fig. \ref{fig:S1}d--e), we computed the two vector components by plugging in the parameters into the solutions in Eqns. (\ref{Eq:GCRM-SS-final}). We plotted the vector components as well as their vector sum (which represents the predicted species and resource sensitivities). For species, we used a single two-dimensional plane representing the changes in abundances of both species. For resources, we split them into occupied (resources 1 and 2) and unoccupied (resource 3, waste) niches. To calculate the prediction error as a function of relative perturbation size, we repeated this analysis by changing the relative size of the environmental perturbations to $m$ and $K$, and calculating the absolute value of the relative error in predicted $\d \vec{N}$.

\comment{\subsection{Comparing competitive and cooperative models}
To simulate the examples in Fig. \ref{fig:5}, we simulated dynamics according to Microbial consumer-resource model. For cooperation, we used the Microbial consumer-resource model with cross-feeding and leakage $\ell=0$, while for competition we used zero leakage, $\ell=0$. It is easily seen that at zero leakage, the model with cross-feeding becomes equivalent to the competitive Linear Resource model with $c_{i\a} = e_{i\a}$. Thus the former models cooperation while the latter models competition. We numerically the cooperative model using the Community Simulator \cite{marsland2020community} using the following parameters: binary sampling, mean consumption rate 10 with standard deviation 5, fraction of secretion flux to the same resource type $f_s=0.9$ and to waste $f_w=0.05$, with external resource supply. We simulated the competitive version using custom numerical code (Community Simulator has a known bug for no leakage $\ell=0$). For the sake of comparison between both models, we used the same randomly generated consumption preferences, mortality rates, and species initial conditions for each simulation. We simulated the model with 100 species and 51 resources (with the last being a waste, unusable resource). In the cooperative model, we supplied only one resource externally at rate $K=100$, while all others were generated through cross-feeding. In the competitive model, all resources except the waste were supplied at equal rates $K_\a=2$. The total resource supply flux was the same for both models. We simulated both models to steady state, and used the steady state abundances to compute the effective species interaction matrices $A_{ij}$ according to the expressions in Eq. (\ref{Eq:AijCrossFeeding}) for cooperation and Eq. (\ref{Eq:AijLinResourceModel}) for competition. These were computed for the surviving species only, which could be different in the case of cooperation and competition, even with the same consumption preferences of all species. Similar results were obtained for different random instantiations. One representative example for each case is shown in Fig. \ref{fig:5}.

\subsection{Bacterial denitrification}
To simulate the examples in Fig. \ref{fig:4}, we simulated dynamics according to the bacterial denitrification model according to Eqns. (\ref{Eq:Denit-dyn}), with two  species and three resources: representing nitrate, nitrite and nitrogen respectively ($S=2$ and $M=3$) and leakage $\ell=0.7$. We set neutral pH to 7, and minimum pH to 1. We supplied only nitrate ($R_1$) at rates $K_1=20$, while the supply for other resources was only via cross-feeding (i.e., $K_2 = K_3 = 0$). Both species had mortality rates $m_1=m_2=1$. The denitrification matrix resembled the chain in Fig. \ref{fig:4}a, with the only non-zero entries being $D_{21}=1$ and $D_{32}=1$. Like in cross-feeding, neither species could consume the waste resource and had consumption preferences $r_{13}=r_{23}=0$. Both species could consume the other two resources with comparable preferences when they were generalists. When specialists, both species instead strongly preferred one of the two resources and only weakly consumed the other (Fig. \ref{fig:S2}a). We obtained similar results with more strict requirements where specialists could only consume one of the two resources (Fig. \ref{fig:S2}b; ``strict'' specialists).  In all cases, the consumption and impact coefficients were identical, i.e., $r_{i\a}=I_{i\a}$. To calculate the simulated and theoretically predicted species sensitivities to changing pH, we followed an analogous procedure to the microbial cross-feeding case in Fig. \ref{fig:4}, where instead of changing the supply and mortality rates of resources and species, we performed a ``global perturbation'' to the pH decreasing it by $\Delta\text{pH}=0.05$, from neutral pH of 7. We simulated the steady states pre and post pH perturbation, and computed the changes in steady-state species abundances. We then calculated the predicted changes using Eqns. (\ref{Eq:Denit-lin}) and (\ref{Eq:Denit-EffPert}), and plotted them against the simulated changes in Fig. \ref{fig:4}d--g (results shown for strict specialists). We repeated this procedure for both generalists and specialists. To gain a statistical understanding of the sensitivities of generalists and specialists, we repeated this simulation procedure to compute the species sensitivities $\d N_1$ and $\d N_2$ for 50 randomly generated sets of generalists and specialists each. In each set, species 1 and 2 had diagonal consumption preferences $r_{11}$ and $r_{22}$ picked as Gaussian random variables with mean 0.6 and standard deviation 0.1. The off-diagonal preferences $r_{12}$ and $r_{21}$ were set to zero mean in the case of specialists and mean 0.6 in the case of generalists, with standard deviation 0.05. These results are plotted in Fig. \ref{fig:S2}.}

\begin{table*}[h]
\centering
\begin{tabularx}{\textwidth}{>{\centering\arraybackslash}X | >{\centering\arraybackslash}X | >{\centering\arraybackslash}X | >{\centering\arraybackslash}X} 
 \hline
 model & species interactions \(A\) & resource interactions \(Q\) & projector \(P\) \\ [0.5ex] 
 \hline\hline
 Generalized consumer-resource model (linear response) & \makecell{\(-B + C^* Q^{-1} E^{* T}\)} & \makecell{\(\delta_{\alpha\beta} + \frac{\partial f_\alpha}{\partial R_\beta}\)} & \makecell{\(E^{* T} A^{-1} C^{*} Q^{-1}\) \\ (only defined when \(B=0\))}\\ 
 \hline
 MacArthur model (\(C=E\)) & \(C^* C^{* T}\) & \(\delta_{\alpha\beta}\) & \(C^{* T} (C^{*}C^{* T})^{-1} C^{*}\) \\ 
 \hline
 MacArthur model (\(C\neq E\)) & \(C^* E^{* T}\) & \(\delta_{\alpha\beta}\) & \(E^{* T} (C^{*}E^{* T})^{-1} C^{*}\)\\ 
 \hline
 Externally supplied resources & \makecell{\(\sum_{\alpha} \frac{g_i \mathrm{w}_{i \alpha} c_{i \alpha} e_{j \alpha} R^*_\alpha}{\omega_\alpha^{\text{eff}}}\)} & \makecell{\(\frac{\omega_\alpha^{\text{eff}}}{\omega} \delta_{\alpha \beta}\)} & \makecell{\( \sum_{jk} \frac{R^*_\alpha e_{j \alpha}[A^{-1}]_{jk}g_k \mathrm{w}_{k \beta} c_{k \beta}}{\omega_\beta^{\text{eff}}}\)}\\
 \hline
 Microbial cross-feeding model & \makecell{\(\omega^{-1} \sum_{\alpha, \beta} \Delta g_{i \alpha} [Q^{-1}]_{\alpha \beta} \kappa_{j \beta}^{\text{eff}}\)} & \makecell{\(\frac{\omega_\alpha^{\text{eff}}}{\omega} \delta_{\alpha \beta}-\Delta \kappa_{\alpha \beta}^{cf}\)} & \makecell{\(-\sum_{jk\gamma} \kappa_{\alpha j}^{\text{eff}} [A^{-1}]_{jk}\Delta g_{k \gamma}[Q^{-1}]_{\gamma \beta}\)}\\
 \hline
  Bacterial denitrification & \makecell{\(\sum_{\alpha, \beta} (1-\ell)r_{i\alpha} [Q^{-1}]_{\alpha \beta} E_{j\beta}\)} & \makecell{\(-\omega - \sum_j I_{j\alpha}N_j^* + \Delta\kappa_{\alpha\beta}^{cf} \)} & \makecell{\(\sum_{jk\gamma} E_{j\alpha} (A^{-1})_{jk} C_{k\gamma} (Q^{-1})_{\gamma\beta} \)}\\
 \hline
\end{tabularx}
\caption{\footnotesize\textsf{\textbf{Effective geometric objects in a variety of consumer-resource models.} Expressions defining the generalized metrics in species space \(A_{ij}\) and resource space \(Q_{\alpha\beta}\), along with the projector \(P^\alpha_\beta\) encoding the metabolic niches occupied by species in the ecosystem, across a variety of consumer-resource models (see Appendix~\ref{Appendix:GCRM-LinResp} for exact definitions of all models).}}
\label{table:geom_objects}
\end{table*}

\end{document}